\documentclass{vldb}
\usepackage{amsfonts}
\usepackage{amsmath,bm}
\usepackage{stmaryrd}
\usepackage{epstopdf}
\usepackage{amssymb}
\usepackage{url}
\usepackage{graphicx}
\usepackage{balance}
\usepackage[tight,footnotesize]{subfigure}
\usepackage{epsfig}
\usepackage[linesnumbered,ruled,vlined]{algorithm2e}
\usepackage{algpseudocode}
\usepackage{caption}
\usepackage{color}
\usepackage{cancel}
\usepackage{mathtools}

\usepackage{tikz}

\DeclareMathOperator*{\argmax}{arg\,max}

\SetKwRepeat{Do}{do}{while}

\usepackage{epstopdf}

\newtheorem{lemma}{Lemma}
\newtheorem{theorem}{Theorem}

\newtheorem{definition}{Definition}

\begin{document}
	
	\title{DTNC: A New Server-side Data Cleansing Framework for Cellular Trajectory Services}
	
	\author{
		\alignauthor
		Jian Dai$^{1}$\hspace{20pt} Fei He$^{1}$  \hspace{20pt}  Wang-Chien Lee$^{2}$\hspace{20pt} Gang Chen$^{3}$\hspace{20pt} Beng Chin Ooi$^{1}$\hspace{20pt}\\
		\affaddr{$^1$School of Computing, National University of Singapore, Singapore}\\
		\affaddr{$^2$Department of Computer Science and Engineering, Pennsylvania State University, USA}\\
		\affaddr{$^3$College of Computer Science, Zhejiang University, China}\\
		\email{\{daij, hef, ooibc\}@comp.nus.edu.sg\hspace{35pt}wlee@cse.psu.edu\hspace{35pt}cg@cs.zju.edu.cn}
	}

	\maketitle
	\begin{abstract}
	It is essential for the cellular network operators to provide cellular location services to meet the needs of their users and mobile applications. However, cellular locations, estimated by network-based methods at the server-side, bear with {\it high spatial errors} and {\it arbitrary missing locations}. Moreover, auxiliary sensor data at the client-side are not available to the operators. 
	In this paper, we study the {\em cellular trajectory cleansing problem} and propose an innovative data cleansing framework, namely \underline{D}ynamic \underline{T}ransportation \underline{N}etwork based \underline{C}leansing (DTNC), to improve the quality of cellular locations delivered in online cellular trajectory services. 
	We maintain a dynamic transportation network (DTN), which associates a network edge with a probabilistic distribution of travel times updated continuously. In addition, we devise an object motion model, namely, {\em travel-time-aware hidden semi-Markov model} ({\em TT-HsMM}), which is used to infer the most probable travelled edge sequences on DTN. 		
	To validate our ideas, we conduct a comprehensive evaluation using real-world cellular data provided by a major cellular network operator		
		and a GPS dataset collected by smartphones as the ground truth.
		In the experiments, DTNC displays significant advantages over six state-of-the-art techniques. 
		\end{abstract}
	
	\vspace{-10pt}
	
	\section{Introduction}

	The market coverage of mobile cellular phones in the world has reached $7$ billion with a penetration rate of $97\%$ by the end of 2015, raised from $12\%$ of the world population in 2000~\cite{itu-int}. 
	%
	%
	As these cellular phones are generating a large amount of cellular data all the time,
	%
	cellular network operators (e.g., StarHub\footnote{A major cellular network operator in Singapore and South-east Asia. See http://www.starhub.com/}) have been seeking ways to utilize such data
	%
	for network maintenance and new service provisioning, e.g., targeted advertisements, location-based services, etc. 
	%
	Indeed, real-time cellular location services,
	which capture the just-in-time human mobility, 
	are long-awaited by not only the location-aware application developers but also some government departments such as the Department of Transportation, 
	for traffic management and urban planning.
	Therefore, it is high-priority for the cellular network operators to provide online cellular location services.
	%
	
\begin{figure}[!htb]
		\centering
		\vspace{-10pt}
		\subfigure[Cellular Trajectory Data Service]{
			\label{fig:illustrative:c} 
			\includegraphics[width=0.47\columnwidth]{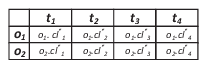}}
		\subfigure[Raw Cellular Trajectories]{
			\label{fig:illustrative:b} 
			\includegraphics[width=0.47\columnwidth]{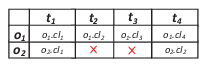}}\\
		\subfigure[A simple transportation network and two raw cellular trajectories]{
			\label{fig:illustrative:a} 
			\includegraphics[width=0.65\columnwidth]{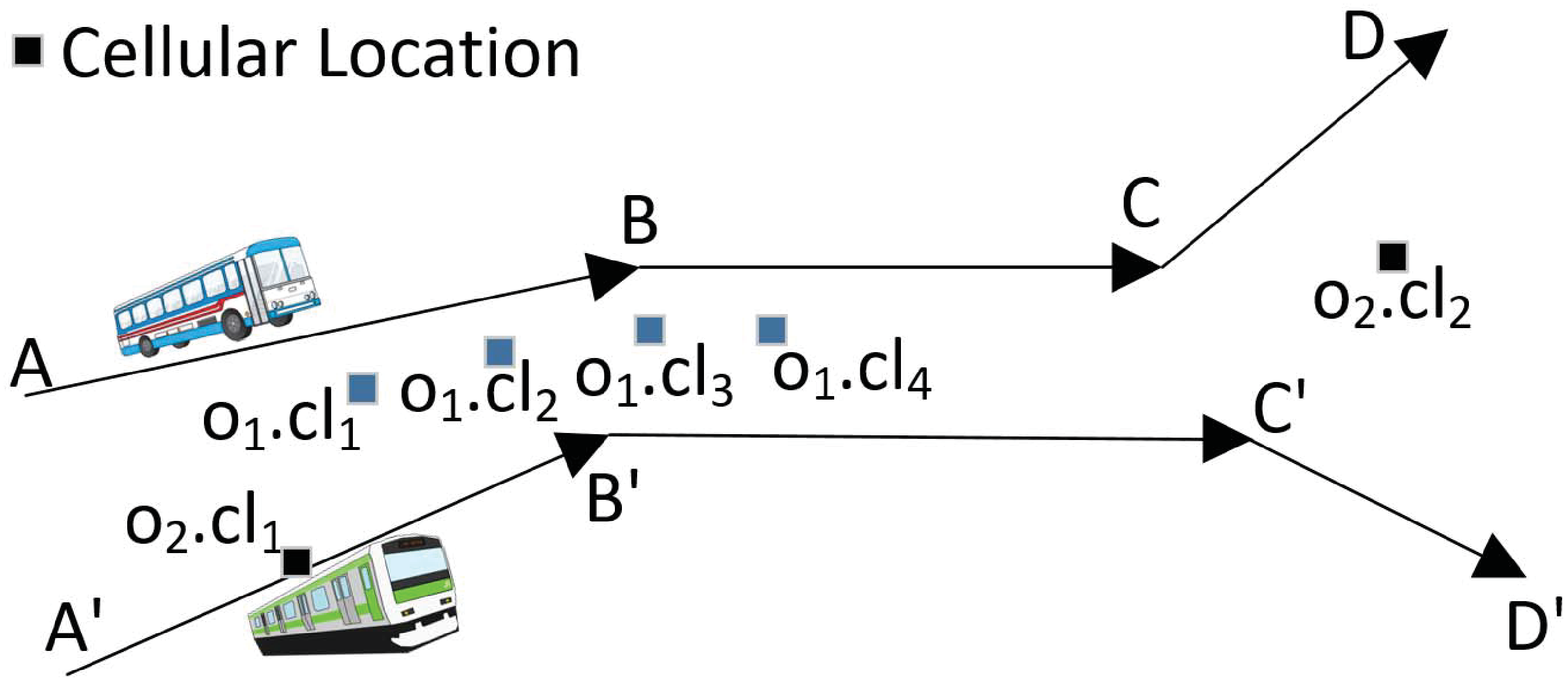}} \vspace{-5pt}
		\caption{Cellular Trajectory Cleansing}
		\vspace{-5pt}
		\label{fig:illustrative} 
	\end{figure}

	Owing to the mobility of cellular users and the tremendously large data volume, cellular location data are often provided in the form of {\em time-windowed trajectories}, which are available timely but only for a period of time due to continuous updates. Figure~\ref{fig:illustrative:c} illustrates a {\em cellular trajectory data service} for two mobile users $o_1$ and $o_2$.\footnote{We refer to the cellular phones/users as {\em mobile objects}.} As shown, the cellular trajectory of an object $o$ within a time window $\mathcal{W}$, denoted by $o.\mathcal{T}^\star$ $=$ $\langle o.cl_1^\star, o.cl_2^\star,$ $\cdots, o.cl_{|\mathcal{W}|}^\star\rangle$, consists of a series of cellular locations, where each cellular location $o.cl=\langle lat,lon,t \rangle$ contains the estimated latitude and longitude of $o$ at time $t$.  

	Cellular trajectory services have great advantages over GPS trajectory services in urban cities where GPS signals may be interfered or blocked by surrounding environments such as buildings and underground public transports.
	%
	In contrast, cellular networks cover nearly all corners of a city. Again, as mentioned earlier, the cellular trajectory data not only benefit location-aware applications but also have potential to support real-time and large-scale analysis on traffic conditions~\cite{calabrese2011estimating,blondel2012data,steenbruggen2015data}, and individual activities~\cite{phithakkitnukoon2010activity,calabrese2013understanding}. 
	%
	
	%
	%
	
	Nevertheless, the poor quality of cellular locations is a major pitfall. Estimated by signal strength, time difference of arrival, and other cellular information~\cite{cong2002hybrid}, cellular locations usually suffer from \textbf{high spatial errors}. As a result, cellular network operators supplement each estimated cellular location with an {\em uncertainty degree} $u$ to indicate its potential spatial error.\footnote{\label{fn:uncertainty-degree} According to StarHub, our cellular network collaborator, the upper bound of spatial error can be estimated by $r(u)=150+50\times(u-1)$, which ranges from $150$~meters ($u=1$) to around $350$~meters ($u=5$).} Additionally, to preserve power in cellular phones, redundant communications between a cellular phone and base stations have been reduced, resulting in \textbf{arbitrary missing locations} in cellular trajectories. Finally, as the cellular locations are estimated by network-based methods at the server-side,  \underline{auxiliary data from sensors on} \underline{smartphones, e.g., GPS, accelerometers, gyroscope, and bar-} \underline{ometer, are not available to the operators.} 

	In this paper, we study the issue of {\em cellular trajectory cleansing (CTC)} for cellular trajectory data services. 
	\begin{definition} [{\bf Cellular Trajectory Cleansing}] Given a set of mobile objects $O$ and a time window $\mathcal{W}=\langle t_1,t_2,$ $\cdots,t_{|\mathcal{W}|} \rangle$, $\forall o \in O, o.\mathcal{T}$ denotes the estimated cellular trajectory of $o$ collected within $\mathcal{W}$. The task of cellular trajectory cleansing is to provide an {\bf\em accurate} and {\bf\em complete} trajectory $o.\mathcal{T}^\star=\langle o.cl_1^\star, o.cl_2^\star, \cdots, o.cl_{|\mathcal{W}|}^\star\rangle$ which consists of $|\mathcal{W}|$ inferred locations corresponding to the times in $\mathcal{W}$. 
	\end{definition}
	
	We use an example to illustrate CTC. Consider two mobile objects $o_1$ and $o_2$ traveling by bus and subway, respectively. The (raw) trajectories estimated by their cellular operator are shown in Figure~\ref{fig:illustrative:b}, where $o_1.\mathcal{T}$ includes four consecutively observed cellular locations, i.e., $o_1.cl_1$, $o_1.cl_2$, $o_1.cl_3$, and $o_1.cl_4$, while $o_2.\mathcal{T}$ includes only $o_2.cl_1$ and $o_2.cl_2$ corresponding to $t_1$ and $t_4$. Obviously, there exist two {\em missing locations} at $t_2$ and $t_3$ in $o_2.\mathcal{T}$. Moreover, as Figure~\ref{fig:illustrative:a} illustrates, all the cellular locations in $o_1.\mathcal{T}$ deviate from the bus route ($A\rightarrow B\rightarrow C\rightarrow D$) where object $o_1$ moves on. Further, $o_2.cl_2$ also deviates from the subway line ($A^\prime\rightarrow B^\prime\rightarrow C^\prime\rightarrow D^\prime$). Thus, CTC aims to transform the noisy raw cellular trajectory data (along with the supplementary spatial error indicators) into {\em accurate} and {\em complete} trajectories as depicted in Figure~\ref{fig:illustrative:c}. 
	
	
	The above example, not only useful for understanding the challenges faced in CTC, offers an idea for cellular trajectory data cleansing. As mobile users are expected to move on roads and other transportation lines, we may exploit a {\em transportation network} (which maps all roads and transport routes in a city) to discover the physical movements of cellular users (on edges of the network) from their raw cellular trajectory data. By calibrating noisy trajectories with the transportation network, we correct spatial errors and infer missing locations to support highly accurate and complete trajectories. 
	
	There exists some relevant research in {\em map matching} which aims to align GPS trajectories with edges on maps (see details in Section 2). However, CTC is different from (and more challenging than) map matching, as CTC aims to infer the exact time-stamped physical locations (i.e., movement) of a mobile object instead of identifying only the network edges where the object travelled. Moreover, map matching algorithms typically assume availability of high-quality GPS trajectory data (and often rely on some auxiliary sensor data). As a result, they do not work well with cellular location data. In this work, we do not only align the noisy cellular trajectories with the network edges but also propose a novel object motion model to effectively deduce the movement on traveled path which in turn infers the cellular locations. 
	
	%
	%
	%
	%
	%
	%
	
	In this work, we propose to build a new data cleansing framework, namely, \underline{D}ynamic \underline{T}ransportation \underline{N}etwork based \underline{C}leansing (DTNC), for cellular trajectory data services. DTNC consists of two major components: (1) a {\em dynamic transportation network (DTN)} that does not only facilitate efficient retrieval of candidate network edges for a cellular location but also provides dynamic travel time information associated with these edges; and (2) a {\em cellular trajectory cleansing} process which takes cellular locations in noisy trajectories to infer the most probable traveled edge sequence (i.e., trajectory on transportation network) and then infer the physical locations of a user for cleansing. At the heart of the trajectory cleansing process is a novel object motion model, namely, {\em Travel-time-aware Hidden semi-Markov Model (TT-HsMM)}, designed to capture the movements of a user on the transportation network, based on her (noisy) cellular trajectory and the travel time information on the DTN. TT-HsMM takes {\em emission probability}, {\em transition probability} and {\em remaining travel duration} of edges estimated from the DTN to differentiate candidate routes, where {\em transition probability} and {\em remaining travel duration} can be computed online with the edge-level travel time distributions. 
	%
	%
	The contributions we made in this research are five-fold:

	\begin{itemize}
		\setlength\itemsep{0em}
		\item We identify the online cellular trajectory cleansing problem, and present a novel cleansing framework, namely DTNC, in support of a cellular trajectory data service, which is essential to many location-aware applications.
		\item We maintain a DTN in DTNC and develop an {\it online travel time distribution learning algorithm} to timely capture the time-dependent travel time information on the DTN.
		It guarantees a theoretical bound for the learned distributions.
		\item We derive a {\em robust emission probability} which represents the likelihood for a cellular location to be generated from a candidate edge fragment. Moreover, we devise an {\em adaptive transition probability} to adaptively depict how a mobile object transits between two edge fragments within a given time period. 
		\item We develop a Travel-time-aware Hidden semi-Markov model (TT-HsMM), to describe the movement of a mobile object on transportation network, based on variable duration and partially observed noisy cellular locations. Moreover, we devise algorithms to infer the most probable edge sequence and physical locations of the object.
		\item We perform extensive experiments with a real-world cellular location dataset and a GPS trajectory dataset (served as ground truth). The results show that DTNC effectively supports CTC by significantly outperforming six relevant methods, namely, PF~\cite{kempinska2016probabilistic}, HMM~\cite{newson2009hidden}, OHMM~\cite{goh2012online}, STRS~\cite{DBLP:conf/kdd/WuMSZZCW16}, SnapNet~\cite{mohamed2016accurate}, and CTrack~\cite{thiagarajan2011accurate}.
		%
	\end{itemize}

	\vspace{-10pt}
	\section{Related Work}
	\label{sec:related}
	
	%
	The following lines of research are relevant to our study.
	
	\noindent\textbf{Transportation Network Modeling}. 
	%
	%
	Basically, two types of information are modeled in a transportation network: (i) static network information such as topology and edge types; (ii) dynamic information such as the travel time and GHG emissions, which are often affected by traffic conditions and hence time-dependent. To answer probabilistic path queries, Hua and Pei~\cite{hua2010probabilistic} propose to associate probabilistic weights with edges in road networks. Yang et al.~\cite{yang2014stochastic} consider the stochastic skyline route planning problem in a road network. A trip planner over probabilistic time-dependent road network is proposed in~\cite{lian2014trip}, which retrieves trip plans traversing a set of query points. In~\cite{dai2016path}, path cost distributions on road network are estimated from historical trajectories. DTN differs from these existing probabilistic road network. First, it models a multimodal transportation network with different edge types. Second, its edge weights are estimated probabilistically by noisy cellular trajectories to reflect the time-dependent travel times on those edges. Third, the edge weights are timely updated with a theoretical bound.
	
	%
	
	\noindent\textbf{Map matching}. 
	As we aim to exploit a DTN to discover the physical movements of cellular users for cellular trajectory cleansing (CTC), map matching algorithms which align GPS trajectories with edges on maps are relevant to this study.
	%
	%
	%
	Most map matching methods~\cite{raymond2012map,newson2009hidden,yin2016general,gustafsson2002particle,wei2012fast,kempinska2016probabilistic} exploit spatial similarity to determine a likely path (i.e., edge sequence under our context) that matches well with a given GPS trajectory. 
	However, due to the high spatial errors in cellular trajectories, the inaccurate spatial similarity may mislead the alignment towards wrong edges. Among those algorithms, Hidden Markov Model (HMM)~\cite{newson2009hidden}, perhaps the most widely used one for GPS data, is proposed for map matching upon noisy and spare trajectory data. 
	Meanwhile, particle filters, previously designed for object tracking, have also been proposed for map matching~\cite{kempinska2016probabilistic,gustafsson2002particle}.
	%
	%
	Thus, we include these two basic methods in the evaluation for comparison.
	
	Some methods employ ``auxiliary information'' besides the observed locations for map matching~\cite{DBLP:journals/sigpro/YuK03,krumm2007map,goh2012online,mohamed2016accurate,hu2017if}.
	For example, in~\cite{DBLP:journals/sigpro/YuK03}, the Hidden semi-Markov Model is applied to combine multiple observation streams from an object to track it.
	OHMM~\cite{goh2012online} relies on the real-time sensor data (i.e., speed and heading direction) on smartphones to derive the speed penalty factor and momentum change, in addition to sensor-deduced traveling distance, to calculate the distance discrepancy between a trajectory and candidate edges for map matching.
	%
	%
	SnapNet~\cite{mohamed2016accurate} employs customized speed filter, $\alpha$-trimmed filter and direction filter before applying an extended HMM. 
	%
	In~\cite{krumm2007map}, travel time constraints, derived from the speed limits, are exploited. 
	Under the context of CTC, sensor data from smartphones are not available. 
	While the time-dependent travel time distributions are incorporated into DTN, they are learned online from the cellular trajectories, without relying on additional auxiliary information.  
	In the evaluation, we include OHMM~\cite{goh2012online} and SnapNet~\cite{mohamed2016accurate} for comparison.
	
	Recently, some research studies how to take low-sampling-rate trajectories as the input to recover the missing traveled edges~\cite{zheng2012reducing,su2015calibrating,DBLP:conf/kdd/WuMSZZCW16}.
	We consider this line of studies as map matching upon trajectories with missing locations.
	Zheng et al. in~\cite{zheng2012reducing} find the candidate traveled edge sequence (i.e., path) according to the reference trajectories.
	In~\cite{su2015calibrating}, trajectories are calibrated via some anchor points, relying on trained inference models from historical trajectory data. 
	In~\cite{DBLP:conf/kdd/WuMSZZCW16}, a Spatio-Temporal-based Route Recovery System (STRS) is presented, which establishes a hybrid model based on high-quality historical trajectories.
	Unfortunately, high-quality cellular trajectories are not available under the context of CTC. Thus, the aforementioned methods may not work effectively on cellular data. 
	%
	%
	In order to validate the proposed ideas in DTNC, we include representative map matching algorithms in our evaluation for comparison. 

	\noindent\textbf{Location inference}.
	Location inference, which derives specific locations from the an inferred edge sequence (i.e., path), is critical to the output quality of \textit{CTC}.
	Most map matching methods~\cite{wu2016only,hu2017if,yin2016general,wei2012fast,goh2012online,lou2009map,newson2009hidden} treat the projection of a GPS point as its physical location, assuming a Gaussian distribution of the measurement error.
	However, 
	cellular locations cleansed based on these methods tend to have high spatial deviation, as indicated by our experiment results. 

	Some indoor localization methods for smartphones or mobile devices~\cite{chen2015fusion,kang2015smartpdr} explore WiFi or bluetooth signal strength fingerprints to improve the localization accuracy.
	However, such client-side sensor data are inaccessible under our context. Thus, these methods are not applicable.
	In CTrack~\cite{thiagarajan2011accurate}, a smoothing-interpolation pipeline is proposed, which requires the auxiliary sensor hints (e.g. six neighboring towers, cell ID, accelerometer, compass, and gyroscope data).
	As the real-time speed, heading direction, and the acceleration can also be approximated by raw cellular trajectory, we also include CTrack~\cite{thiagarajan2011accurate} in our evaluation for comparison. 
	
	\section{The DTNC Framework}
	\label{sec:dtn-octc}
	
	We first introduce the core concepts in the proposed DTNC and then provide an overview of the framework.
	
	\vspace{-5pt}
	\subsection{Core Concepts}
	
	
	As discussed earlier, {\em dynamic transportation network (DTN)} plays an important role in our framework as it maintains two key information: 1) network structure; 2) accurate and timely captured travel times with respect to edges. The DTN is formally defined as follows.
	
	\begin{definition}[\textbf{Dynamic Transportation Network}]
		\label{def:dtn}
		A dynamic transportation network $DTN=\{G,M\}$ consists of a graph $G$ denoting the structure of transportation network and a mapping $M$ associating traffic information with graph edges. Here, $G=(V,E)$ is a directed graph; $V$ is a vertex set; and $E\subseteq V\times V$ is an edge set.  
		A vertex $v_i\in V$ models an intersection of edges or an edge end.
		$e_k=(eid,v_i,$ $v_j)\in E$ represents a directed edge with  $eid$, indicating that traveling from start vertex $v_i$ (also denoted by $e_k.s$) to end vertex $v_j$ (also denoted by $e_k.d$) is feasible. 
		Each $e_k$ has an edge type, e.g., trunk, motorway, subway, footway, etc.
		The mapping set $M=\{\langle e_1,\mathcal{D}_{e_1}\rangle, \cdots, \langle e_n,$ $\mathcal{D}_{e_n}\rangle\}$ associates a travel time distribution $\mathcal{D}_{e_i}$ with the transportation edge $e_i \ (\forall e_i \in E)$.
	\end{definition}
	
	In a DTN, $G$ models the network structure, and $M$ captures the travel time distributions associated with edges.
	Note that the traffic conditions within a transportation network are time-dependent~\cite{chang2005multiobjective,yang2014stochastic,DBLP:conf/kdd/WuMSZZCW16} for most edges.
	Therefore, the $\mathcal{D}_{e} \ (\forall e \in E)$ is updated in a real-time fashion.
	
	
	We intend to capture cellular locations (movement) of a mobile object on transportation network by an object motion model (to be detailed in Section~\ref{sub:sec:state-duration-prob}), which requires two pieces of information:
	1) emission probability $p(cl_k|x_k)$, which depicts how each cellular location $cl_k$ is generated from a state $x_k$; 2) transition probability $p(x_k|x_{k-1})$, which describes how the underlying movement takes place between two states $x_{k-1}$ and $x_{k}$.
	%
	We argue it's important to use a ``robust'' $p(cl_k|x_k)$ to address the inherent {\em high spatial error} and to derive an ``adaptive'' $p(x_k|x_{k-1})$ based on real-time travel times to tackle the issue of {\em arbitrary missing locations}. To this end, we define {\em robust emission probability}, which takes a transportation {\em edge fragment} as a whole, to consider how likely an observed cellular location is emitted from such an edge fragment.\footnote{An {\em edge fragment} $ef=(eid,p_l,p_r)$ is a part (from $p_l$ to $p_r$) of a candidate transportation edge $e$, where $e.eid=ef.eid$, and $p_l$, $p_r$ are two points on $e$.}
	We also define  {\em adaptive transition probability}, conditioning on not only the previous state but also the time interval between the two states on the DTN at a given time, to determine the likelihood of a state transition.
	
	\begin{definition}[\textbf{Robust Emission Probability}]
		For a cellular location $cl$, the robust emission probability $p(cl|ef\in \mathcal{R}_{cl})$ is the probability for the physical (true) location of $cl$ to reside in a given edge fragment $ef\in\mathcal{R}_{cl}$, where $\mathcal{R}_{cl}$ consists of all the possible edge fragments corresponding to $cl$.
	\end{definition}
	
	Recall that the uncertainty degree $u$ associated with $cl$ gives an upper bound of spatial error between $cl$ and its physical location. We may issue a range query $RQ_{ef}(cl,G)$, using $cl$ as the root and distance derived from its $u$ as the radius, to retrieve $\mathcal{R}_{cl}$, the set of all edge fragments 
	fallen within the region. 
	
	\begin{definition}[\textbf{Adaptive Transition Probability}]
		For two consecutive cellular locations $cl_k$ and $cl_{k+1}$, the adaptive transition probability $p(ef_{k+1}|ef_k,\Delta_k t)$ is the probability for the actual transition from $ef_{k}$ to $ef_{k+1}$ within $\Delta_k t$, where $\Delta_k t=cl_{k+1}.t-cl_{k}.t$ is the period of transition.
	\end{definition}
	
	Instead of a physical location, we model the {\em robust emission probability} on an edge fragment, making it reliable. Moreover, the enrichment of the dynamic conditions
	on DTN brings fine-grained and precise factors into {\em adaptive transition probability}, rendering it adaptive to traffic dynamics and thus more accurate. 
	
	\begin{figure}[!htb]
		\centering
		\vspace{-10pt}
		\subfigure[$G\in DTN$]{
			\label{fig:hsmm:g} 
			\includegraphics[width=0.48\columnwidth]{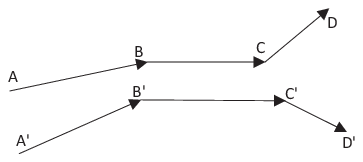}}
		\subfigure[$M\in DTN$]{
			\label{fig:hsmm:m} 
			\includegraphics[width=0.48\columnwidth]{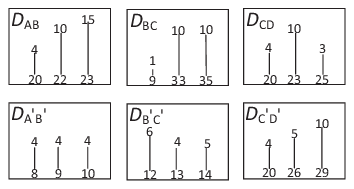}}\\
		\subfigure[$p(cl|ef\in\mathcal{R})$]{
			\label{fig:hsmm:em} 
			\includegraphics[width=0.55\columnwidth]{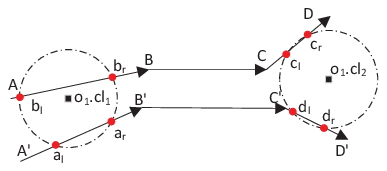}}
		\subfigure[$p(ef_{k+1}|ef_k,\Delta_k t)$]{
			\label{fig:hsmm:tran} 
			\includegraphics[width=0.4\columnwidth]{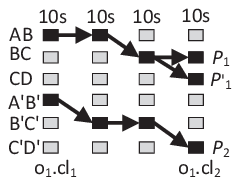}}
		\vspace{-5pt}
		\caption{Illustration of Core Concepts}
		\vspace{-5pt}
		\label{fig:hsmm}  
	\end{figure}
	
	Figure~\ref{fig:hsmm} illustrates the ideas behind these concepts via a simple example, which shows how the dynamically maintained travel time distributions in $DTN$, the {\em robust emission probability} and the {\em adaptive transition probability} contribute to the selection of a sound path via the proposed object motion model. 
	A toy example of the network structure $G\in DTN$ is shown in Figure~\ref{fig:hsmm:g}, while the association of travel time distributions and edges $M\in DTN$ are shown in Figure~\ref{fig:hsmm:m}.
	We use a bar chart (a.k.a. line graph) to represent a travel time distribution of an edge, where each bar is associated with two numbers, standing for the travel time and frequency, respectively.
	For instance, $\mathcal{D}_{BC}$ has three bars, denoting the travel times $9$, $33$, and $35$ seconds (beneath the bars) and the corresponding frequencies $1$, $10$, and $10$ (above the bars), respectively.

	Figure~\ref{fig:hsmm:em} illustrates the {\em robust emission probabilities} w.r.t. $o_1.cl_1$ and $o_1.cl_2$.
	The circular regions centered at $o_1.cl_1$ and $o_1.cl_2$ bound the possible physical locations of them according to $o_1.cl_1.u$ and $o_1.cl_2.u$.
	As a result, we obtain $\mathcal{R}_{o_1.cl_1}=\{ef_a=\langle A^\prime B^\prime.eid, a_l, a_r \rangle, ef_b=\langle AB.eid,$ $b_l, b_r \rangle \}$.
	Accordingly, we consider two {\em robust emission probabilities} w.r.t. $o_1.cl_1$, including $p(o_1.cl_1|ef_a)$ and $p(o_1.cl_1|ef_b)$.
	Similarly, two {\em robust emission probabilities} w.r.t. $o_1.cl_2$ are considered: $p(o_1.cl_2|ef_c)$ and $p(o_1.cl_2|ef_d)$, where $ef_c=\langle $ $C D.eid, c_l, c_r \rangle$ and $ef_d=\langle C^\prime D^\prime.eid, d_l, d_r \rangle$.
	
	Figure~\ref{fig:hsmm:tran} illustrates the {\em adaptive transition probabilities} through a trellis diagram, where each column corresponds to a time slice (of $10$ seconds in this diagram) and each row corresponds to an edge.
	Suppose $\Delta t=o_1.cl_2.t-o_1.cl_1.t=30$~s.
	In this case, observations for the middle two time slices are missing.
	Bearing with missing observations, we compare three candidate paths (i.e., sequences of edges), $\mathcal{P}_1$, $\mathcal{P}_1^\prime$ and $\mathcal{P}_2$, based on their possible internal transitions.
	If the mobile object $o_1$ starts from a position on $ef_b$, as indicated by $\mathcal{P}_1$, $o_1$ should very likely stop at somewhere on $BC$, as suggested by $\mathcal{D}_{AB}$ and $\mathcal{D}_{BC}$.
	While the internal transition of $\mathcal{P}_1$ has a high probability, $\mathcal{P}_1$ is invalid since it does not stop at $ef_c$.
	$\mathcal{P}_1^\prime$ owns the same first three latent states with $\mathcal{P}_1$, except that it uses $CD$ as its last state.
	According to $\mathcal{D}_{AB}$, $\mathcal{D}_{BC}$ and $\mathcal{D}_{CD}$, there exists a small probability for such a transition to happen, because there is a bar in $\mathcal{D}_{BC}$ with travel time $9$~seconds.
	Note that such a transition is an event with small probability since the corresponding frequency is $1$ while the rest two frequencies are $10$. 
	In contrast, if $o_1$ starts from a position on $ef_a$, as suggested by $\mathcal{P}_2$, $o_1$ is able to reach somewhere within $C^\prime D^\prime$, as implied by $\mathcal{D}_{A^\prime B^\prime}$, $\mathcal{D}_{B^\prime C^\prime}$ and $\mathcal{D}_{C^\prime D^\prime}$.
	As $\mathcal{P}_2$ ends at a place within $ef_d$, it is easy to accept that $\mathcal{P}_2$ is a sound path with high internal transition probability, according to the $\Delta t$ and $DTN$. 

	\subsection{Framework Architecture}

\begin{figure}[!htb]
		\centering
		\vspace{-10pt}
		\includegraphics[width=0.95\columnwidth]{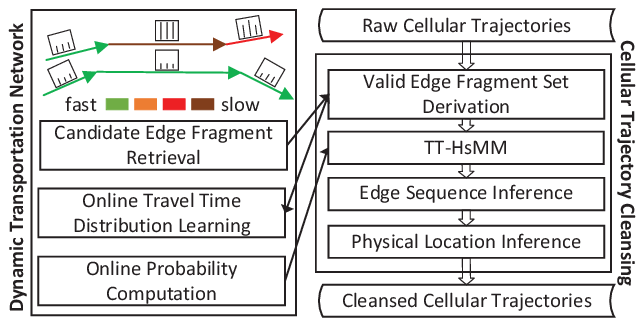}
		\caption{Overview of the DTNC framework}
		\label{fig:framework} 
	\end{figure}

	Based on the core concepts, we describe the proposed DTNC framework for cellular trajectory cleansing. Figure~\ref{fig:framework} gives an overview of the DTNC framework which consists of two major components: (1) {\em Dynamic Transportation Network} and (2) {\em Cellular Trajectory Cleansing}.
	
	\noindent{\bf Dynamic Transportation Network. }According to  Definition~\ref{def:dtn}, this component 
	%
	maintains the static network structure and dynamic traffic information in a DTN to facilitate: 1) {\em candidate edge fragment retrieval}
	- issuing a range query to retrieve candidate edge fragments for a cellular location; 
	and 2) {\em online probability computation} - computing {\em robust emission probability} and {\em adaptive transition probability}. 
	In addition, DTN is updated by 3) 
	{\em online travel time distribution learning} 
	- learning the travel time distribution from just-in-time updates to capture the traffic changes. 
	
	\noindent{\bf Cellular Trajectory Cleansing. }This component realizes the process of data cleansing in four steps: 1) {\em valid edge fragment set derivation} - as noisy raw cellular trajectories stream in, it invokes candidate edge retrieval and further derives valid candidate edge fragments; 
	2) {\em TT-HsMM} - the object motion model is constructed online based on the valid edge fragments, leveraging the {\em robust emission probability}, {\em adaptive transition probability}, and the {\em remaining travel duration} of edges estimated by the DTN;
	3) {\em edge sequence inference} - it infers the most probable traveled edge sequence based on the TT-HsMM; and 4) {\em physical location inference} - it infers the physical location w.r.t. each time stamp in the service window. 
	
	\section{Dynamic Transportation Network}
	\label{sec:pro-com} 
	
	Intrinsically, dynamic transportation network serves as an evolving knowledge base, 
	%
	which offers three functionalities:
	%
	1) it supports efficient range query processing to obtain  $\mathcal{R}_{cl}$ for a cellular location;
	2) it refreshes the $\mathcal{D}_{e}$ $(\forall e\in E)$ via online learning;
	3) it computes the probabilities required by the trajectory cleansing component.
	
	%
	
	\subsection{Candidate Edge Fragment Retrieval}
	\label{sub-sec:eff-rcl-retrieval}
	
	Let $\mathcal{R}_{cl}$ denote the set of edge fragments $cl$ may possibly reside.
	%
	We issue a region query $RQ_{ef}(cl,G)$ over $G\in DTN$ to get the edge fragments within a circular region centered at $cl$ with radius $r$ derived from the uncertainty degree $u$ of $cl$ (cf. footnote~\ref{fn:uncertainty-degree}).
	Note that the computation of $r$ involves the definition of an uncertainty model for cellular locations, which is orthogonal to $DTNC$.
	%
	Other uncertainty models (e.g. ellipse error model~\cite{pfoser1999capturing}) can be used as well.
		
	\subsection{Online Travel Time Distribution Learning }
	\label{subsec:de}
		
	Here we describe how to learn and update travel time distribution $\mathcal{D}_e$ for edge $e$ on DTN.   
	
	\textbf{Initial $\mathcal{D}_e$}. We initialize 
	$\mathcal{D}_e$ 
	by the travel speed limit $d_e$ of edge $e$. Specifically, a collection of travel speed values $\{s_1, s_2,\cdots, s_n\}$ are sampled from the Uniform distribution $\mathcal{U}(0,d_e)$. Thus, $\{\frac{|e|}{s_1}, \frac{|e|}{s_2}, \cdots, \frac{|e|}{s_n}\}$, where $|e|$ denotes the length of $e$, forms the discrete travel time distribution $\mathcal{D}_e$.
	
	As new cellular trajectories constantly arrive in DTNC, we exploit them to compute online the travel time on edge $e$ and update the distribution $\mathcal{D}_e$. Instead of using all the edge fragments derived from cellular locations to compute the travel time on edges, we consider only the {\em most compact edge fragment sets} because they contain only one, most certain edge fragment.\footnote{After the edge fragment set $\mathcal{R}_{cl}$ for an edge $e$ is retreived, invalid edge fragments are filtered (detailed in Section~\ref{sub-sec:break}). The valid edge fragment set with one remaining edge fragment, denoted by $\mathcal{R}_{cl}^\star$ is the most compact and certain.}   
	
	\textbf{Online travel time computation}. 
	Consider two most compact edge fragment sets $\mathcal{R}_{cl_i}^\star=\{ef_\alpha\}$ and $\mathcal{R}_{cl_{i+1}}^\star=\{ef_\beta\}$, corresponding to two consecutive cellular locations, $cl_i$ and $cl_{i+1}$.
	If $ef_\alpha$ and $ef_\beta$ belong to the same edge $e$, the travel time of $e$ can be estimated as follows.
	First, the travel speed $s_e$ for $e$ is computed by $\frac{|ef_\alpha.p_c, ef_\beta.p_c|}{cl_{i+1}.t-cl_i.t}$, where $ef_\alpha.p_c$ and $ef_\beta.p_c$ denote the midpoint of $ef_\alpha$ and $ef_\beta$, respectively.
	Then, the travel time $t_e$ of the whole edge $e$ is computed by $t_e=\frac{|e|}{s_e}$, where $|e|$ denotes the length of $e$.

	\textbf{Update $\mathcal{D}_e$.} 
	Within the current time window $\mathcal{W}$, a collection of travel times $\{t_e^1, t_e^2, \cdots, t_e^n\}$ corresponding to an edge $e$ may be computed over the most compact edge fragment sets. 
	Accordingly, we update the $\mathcal{D}_e$ based on the \textit{Hoeffding bound} (a.k.a. \textit{additive Chernoff bound})~\cite{hoeffding1963probability}. 
	
	Let $\mathcal{T}_e$ be the random variable described by $\mathcal{D}_e$,  $\widehat{t_e}$ be the arithmetic mean of $\{t_e^1, t_e^2, \cdots, t_e^n\}$ and $R$ be the range of all outcomes of $\mathcal{T}_e$. 
	Initially, $R$ is assigned to the value computed from $\mathcal{D}_e$ before update. 
	After the arrival of $\{t_e^1, t_e^2, \cdots, t_e^n\}$, Hoeffding bound states that the deviation between the empirical mean $\widehat{t_e}$ and the actual mean $E(\widehat{\mathcal{T}_e})$ is at most $\epsilon$ with the probability $1-\delta$, where  $\epsilon=\sqrt{\frac{R^2ln(1/\delta)}{2n}}$. 
	
	%
	According to this bound, if the time interval corresponding to $\mathcal{D}_e$ is narrowed (i.e., $R$ is reduced), $\epsilon$ drops as long as the $n$ remains unchanged (since $\delta$ is predefined).  
	Nevertheless, the reduction in $R$ leads to the decrease of $n$ as well, rendering the narrowing of $R$ a subtle task.
	
	We aim to reduce $R$ with a minor decrease of $\epsilon$, i.e., without sacrificing much accuracy. To this end, we propose a \textit{narrowing} process which identifies the proper endpoints from the previous loose interval and forms a more compact interval iteratively. 
	Let $\delta$ and $\epsilon$ be the predefined thresholds. For instance, $\delta=0.05$ and $\epsilon=1$, meaning that we aim to narrow the previous interval to an extent so that the deviation between the empirical mean $\widehat{t_e}$ of the new interval and the expectation of the actual mean $E(\widehat{\mathcal{T}_e})$ is at most $1$ time slice with the probability $1-0.05=0.95$. This indicates an insignificant accuracy loss.
	
	Given $\delta$ and $\epsilon$, $R=\sqrt{\frac{2n\times \epsilon^2}{ln(1/\delta)}}$ is derived from the original Hoeffding bound  $\epsilon=\sqrt{\frac{R^2ln(1/\delta)}{2n}}$. Since $n$ is also a constant, $R$ can be directly computed. For instance, if $\delta=0.05$, $\epsilon=1$ and $n=20$, $R=\sqrt{\frac{2\times 20\times 1^2}{ln(1/0.05)}}\approx13.35$. 
	%
		
	We describe how to iteratively compute the endpoints of the compact interval. 
	Let $t_l$ and $t_r$ be the left and right endpoints of the previous interval $I$. One of them is to be removed in current iteration. Since a compact interval is desired, the one that can narrow $I$ further is firstly considered. 
	The narrowing ability is evaluated by the distance between it and its adjacent travel time value. Let $t_{l^\prime}$ and $t_{r^\prime}$ be the first values that are close to $t_l$ and $t_r$, respectively. If $|t_{l^\prime}-t_l|>|t_{r}-t_{r^\prime}|$ and $|t_{r}-t_{l^\prime}|>R$, $t_l$ is removed. Otherwise, if $|t_{l^\prime}-t_l|\leq|t_{r}-t_{r^\prime}|$ and $|t_{r^\prime}-t_{l}|>R$, $t_r$ is removed. If both endpoints are unsuitable to be removed (if removed, the computed $R$ will be violated), the interval is kept and the \textit{narrowing} process ends.
	After the removal, a new interval $I^\prime=[t_l^\prime, t_r^\prime]$ is generated. Clearly, if $t_r$ is removed, $[t_l^\prime, t_r^\prime]=[t_l, t_{r^\prime}]$. Otherwise,
	$[t_l^\prime, t_r^\prime]=[t_{l^\prime}, t_{r}]$. 
	The removal loops by using the new interval as the input.
	Note that each iteration removes one endpoint, indicating the new interval contains less travel time values (smaller $n$).
	Hence, the removal loop (i.e., {\it narrowing process}) will eventually stops because $\sqrt{\frac{2n\times \epsilon^2}{ln(1/\delta)}}$ monotonically decreases with $n$.
		
	\begin{figure*}[!htbp]
		\centering
		\subfigure[Update $\mathcal{D}_e$]{
			\label{fig:de-update}  
			\includegraphics[width=0.75\columnwidth]{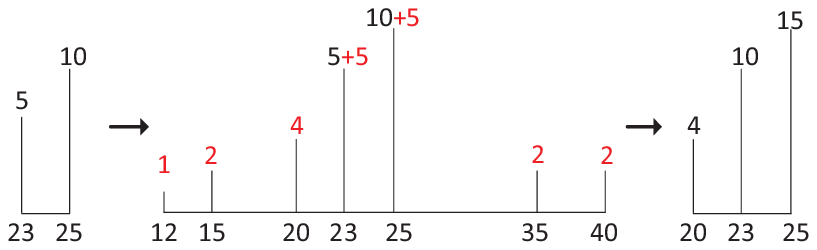}}
		\subfigure[\textit{Narrowing} Process]{
			\label{fig:narrow-example} 
			\includegraphics[width=1.25\columnwidth]{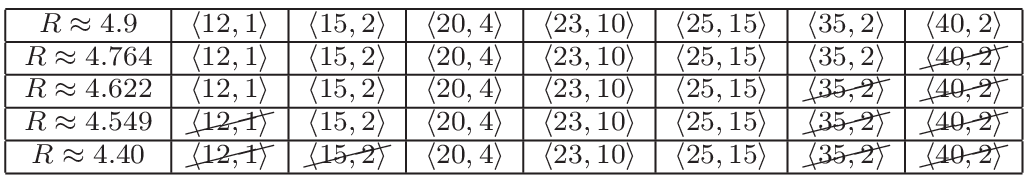}}
		\vspace{-5pt}
		\caption{Online $\mathcal{D}_e$ Learning}
		\label{fig:online-learn}  
		\vspace{-5pt}
	\end{figure*}
	
	After obtaining the compact interval $I^\prime=[t_l^\prime, t_r^\prime]$, the updated distribution $\mathcal{D}_e^\prime$ is calculated based on the frequency of involved travel times in the $I^\prime$. A tuple $\langle t, p_{t}\rangle$ is maintained in $\mathcal{D}_e^\prime$ for each $t$ in $I^\prime$, where $t$ is an integer value representing the travel time and $p_{t}$ is the probability of the travel time $t$ for edge $e$. 
	Note that the computed travel time $t_e^i$ in $\{t_e^1, t_e^2, \cdots, t_e^n\}$ can be a real value, we turn $t_e^i$ into an integer by a ceiling function $ceil(t_e^i)=\lceil t_e^i\rceil$. 
	As such, $p_{t}$ is computed by $p_{t}=p(t|\mathcal{D}_e)=\frac{N_{t}}{N_{I^\prime}}$, where $N_{t}$ is the number of travel times whose ceilings are $t$ and $N_{I^\prime}$ is the number of travel times whose ceilings fall into $I^\prime$.
	
	Suppose a $\mathcal{D}_e$ contains a collection of travel time tuples (where the first element is the travel time of $e$, and the second one is the observed count): $\{\langle 23, 5\rangle$, $\langle 25, 10\rangle\}$ and a collection of new travel time tuples $\{\langle 12,1\rangle$, $\langle 15,2\rangle$, $\langle 20,4\rangle$, $\langle 23,10\rangle$, $\langle 25,15\rangle$, $\langle 35,2\rangle$, $\langle 40,2\rangle\}$ are received, Figure~\ref{fig:de-update} exemplifies how $\mathcal{D}_e$ is updated. 
	Let $\epsilon=1$, $\delta=0.05$. Thus,\\
	$R=\sqrt{\frac{2n\times \epsilon^2}{ln(1/\delta)}}=\sqrt{\frac{2\times (1+2+4+10+15+2+2)\times 1^2}{ln(1/0.05)}}\approx 4.9\ll 28=40-12$, indicating the initial time interval $[12,40]$ is too loose. The first row in Figure~\ref{fig:narrow-example} shows the initial travel times and computed $R$.
	The \textit{narrowing} process removes $\langle 40,2\rangle$ in the first iteration for $40-35>15-12$ and the removal of $\langle 40,2\rangle$ does not violate the newly computed $R=\sqrt{\frac{2\times (1+2+4+10+15+2)\times 1^2}{ln(1/0.05)}}\approx 4.764$ for $4.764<35-12$.
	The \textit{narrowing} process continues until it faces the remaining travel time tuples $\{\langle 20,4\rangle$, $\langle 23,10\rangle$, $\langle 25,15\rangle\}$. In this iteration, the removal of either $\langle 20, 4\rangle$ or $\langle 25,15\rangle$ generates an interval that violates the corresponding $R$. Thus, the \textit{narrowing} process stops.
	
	Based on $\{\langle 20,4\rangle$, $\langle 23,10\rangle$, $\langle 25,15\rangle\}$, a travel cost distribution is computed $p(t=20|\mathcal{D}_e)=\frac{4}{4+10+15}\approx 0.138$,  $p(t=23|\mathcal{D}_e)\approx 0.345$, and $p(t=25|\mathcal{D}_e)\approx 0.517$. Further, $E(\widehat{\mathcal{T}_{e}})\approx \frac{20\times 4+23\times 10+25\times 15}{4+10+15}\approx 23.62$ seconds.
	
	\subsection{Online Probability Computation}
	\label{sec:prob-comp}
	
	Next, we describe how to compute robust emission probability and adaptive transition probability which are needed in the proposed object motion model for edge sequence inference (see Section 5.2-5.3).
	
	\subsubsection{Robust Emission Probability}
	\label{sec:robust-adaptive}
	
	To compute the robust emission probability w.r.t. a cellular location $cl$ and an edge fragment $ef$, i.e., $p(cl|ef\in \mathcal{R}_{cl})$, we follow the maximum entropy principle to assume that any physical location on an edge fragment $ef\in \mathcal{R}_{cl}$ has the same probability $\kappa$ of generating $cl$.
	
	Since each edge fragment $( eid,p_l,p_r )$ is a line segment, the probability $p(cl|ef\in \mathcal{R}_{cl})$ can be obtained by integrating over the line segment. 
	%
	We have the following lemma.
	
	\begin{lemma}
		\label{le:p-em-1}
		$p(cl|ef\in \mathcal{R}_{cl})\propto |ef|$
	\end{lemma}
	
	\textit{Proof: } 
	$p(cl|ef\in \mathcal{R}_{cl})$
	$=\int_{p_l}^{p_r} \kappa dp=|ef|\times\kappa\propto |ef|~~~~~~~\square$
	
	Based on Lemma~\ref{le:p-em-1}, we know that a {\em robust emission probability} $p(cl|ef\in \mathcal{R}_{cl})$ is proportional to the length of $ef$.
	In fact, all possible edge fragments are in $\mathcal{R}_{cl}$.
	Thus, the sum of all the {\em robust emission probabilities} w.r.t. a cellular location is $1$.
	Accordingly, each {\em robust emission probability} w.r.t. $cl$ can be computed, as shown in Lemma~\ref{le:p-em-2}.

	\begin{lemma}
		\label{le:p-em-2}
		$p(cl|ef_j\in \mathcal{R}_{cl})=\frac{|ef_{j}\in\mathcal{R}_{cl}|}{\sum_{ef_i\in \mathcal{R}_{cl}}|ef_i|}$
	\end{lemma}
	
	\textit{Proof: }
	Since $\sum_{ef_i\in \mathcal{R}_{cl}} p(cl|ef_i)=1$, \\
	and let $\mathcal{R}_{cl}=\{ef_1,\cdots,ef_i,\cdots,ef_{|\mathcal{R}_{cl}|}\}$, \\
	$p(cl|ef_1):\cdots :p(cl|ef_i): \cdots p(cl|ef_{|\mathcal{R}_{cl}|}) =|ef_1|:\cdots :|ef_i|:\cdots :|ef_{|\mathcal{R}_{cl}|}|$ (due to Lemma~\ref{le:p-em-1}).\\
	~~~~Therefore, $p(cl|ef_j\in \mathcal{R}_{cl})=\frac{|ef_{j}\in \mathcal{R}_{cl}|}{\sum_{ef_i\in \mathcal{R}_{cl}}|ef_i|}~~~~~~~~~~~~~~~~~~~~~~~~~~~~\square$
		
	\subsubsection{Adaptive Transition Probability}
	\label{sub-sec:hsmm}
	
	Given two consecutive cellular locations $cl_k$ and $cl_{k+1}$, we  approximate 
	$p(ef_{j}\in\mathcal{R}_{cl_{k+1}}|ef_i\in\mathcal{R}_{cl_k},\Delta_k t)$  
	(i.e., the adaptive transition probability) as follows.
	
	We view the adaptive transition probability as a posterior probability, and use Monte Carlo simulations over the $DTN$ to approximate its expectation.
	
	
	For each $ef_i\in\mathcal{R}_{cl_k}$, such a simulation is accomplished in three steps.
	First, the particles in a particle set  $\mathcal{A}=\{par_1, par_2$, $\cdots, par_{|\mathcal{A}|}\}$ are evenly placed within $ef_i$.
	Second, each particle moves along current edge $e$, where $e.eid=ef_i.eid$. 
	When the travel time $t_e=t^\prime_e\times\frac{|par.l,e.d|}{|e|} $ is used up, the particle randomly transits to another edge $e^\prime$, and $\Delta_k t=\Delta_k t-t_e$.
	%
	Here, $t^\prime_e$ is sampled from $\mathcal{D}_e$, $e^\prime.s=e.d$, $par.l$ denotes the particle's location, and $|par.l,e.d|$ represents the distance between $par.l$ and $e.d$.
	Third, the second step repeats itself, until the overall required time $\Delta_k t$ elapses. 
	Then, the particle stops at a place.
	An indicator function $\mathbb{I}(par.l\in ef_{j})$ is devised to examine if the place lies in the edge fragment $ef_{j}$.
	$\mathbb{I}(par.l\in ef_{j})$ returns $1$  if a particle stops at a place within $ef_{j}$; otherwise, it returns $0$.
	In this way, 
	
	$E(ef_{j}|ef_i,\Delta_k t)\approx \frac{ \sum_{par\in\mathcal{A}}\mathbb{I}(par.l\in ef_{j})}{\sum_{ef_n\in \mathcal{R}_{cl_{k+1}}}\sum_{par\in\mathcal{A}}\mathbb{I}(par.l\in ef_n)}$.
	\\
	
	Here, $\mathcal{R}_{cl_{k+1}}$ covers all possible edge fragments for $cl_{k+1}$ and $par\in\mathcal{A}$ covers all the particles in the particle set $\mathcal{A}$. 

	As a genetic method to compute the adaptive transition probabilities for any two edge fragments $ef_i\in \mathcal{R}_{cl_{k}}$ and $ef_j\in \mathcal{R}_{cl_{k+1}}$, we observe that when $\Delta_k t$ is large, the diffused particles after simulations can be sparse, and some edge fragments in $\mathcal{R}_{cl_{k+1}}$ have no particle. 
	Thus, the probabilities associated with them are zeros, which can be inaccurate. 

	Surely, adding more particles can put the sparseness to rest.
	%
	Nevertheless, it brings a computational burden.
	Instead, to fix such sparseness problem with moderate particles, we add a Dirichlet prior $Dir(\bm{\gamma})$ to the distribution of the transited particles. That is,
	
	\noindent$\mathbf{A}_i^k\sim Dir(\gamma_0m_1, \cdots, \gamma_0m_{|\mathcal{R}_{cl_{k+1}}|})=Dir(\gamma_0\mathbf{m})=Dir(\bm{\gamma})$

	\noindent where $\mathbf{A}_i^k$ is the adaptive transition probability vector 
	regarding to the edge fragment $ef_i\in \mathcal{R}_{cl_k}$,
	$\mathbf{m}$ is the prior mean vector (satisfying $\sum_{j=1}^{|\mathcal{R}_{cl_{k+1}}|} m_j=1$, where $m_j$ corresponds to $ef_j\in \mathcal{R}_{cl_{k+1}}$), 
	$\gamma_0$ is the prior strength. 
	
	Accordingly, after obtaining the Multinomial samples from the particles, the posterior $\mathbf{A}_i^k\sim Dir(\bm{\gamma}+\bm{N_{i}})$, where $\bm{N_i}=(N_{i,1}, \cdots, N_{i,|\mathcal{R}_{cl_{k+1}}|})$ is the vector that records the numbers of the particles traveling from the edge fragment $ef_i\in \mathcal{R}_{cl_k}$ to all edge fragments in $\mathcal{R}_{cl_{k+1}}$.
	Specifically, $N_{i,j}$ denotes the number of particles that have traveled from $ef_i\in \mathcal{R}_{cl_k}$ and successfully stopped at $ef_j\in \mathcal{R}_{cl_{k+1}}$ after $\Delta_k t$, and $N_i=\sum_{j=1}^{|\mathcal{R}_{cl_{k+1}}|} N_{i,j}$.
	As such, we can approximate the {\em adaptive transition probability} as follows.
	
	\begin{lemma}
		\label{le:p-atran}
		$p(ef_{j}|ef_i,\Delta_k t)\approx\frac{N_{i,j}+\gamma_0 m_j}{N_i+\gamma_0}$
	\end{lemma}
	
	\textit{Proof: } 
	$p(ef_{j}|ef_i,\Delta_k t)\approx\frac{N_{i,j}+\gamma_0 m_j}{\sum_{n=1}^{|\mathcal{R}_{cl_{k+1}}|} (N_{i,n}+\gamma_0 m_n)}$\\
	$~~~~~~~~~~~~~~~~~~~~~~~~~~~~~~~~~~~~~~=\frac{N_{i,j}+\gamma_0 m_i}{N_i+\gamma_0}~~~~~~~~~~~~~~~~~~~~~~~~~~~~~~~\square$
	
	
	\section{Cellular Trajectory Cleansing}
	\label{sec:cell-clean}
	
	For the raw cellular trajectories within current time window $\mathcal{W}$, 
	we first perform {\em valid edge fragment set derivation}, 
	to derive valid (and more compact) edge fragments;
	then we instantiate TT-HsMM model to infer the most probable edge sequence; 
	lastly, we infer the physical locations. 
		
	\subsection{Valid Edge Fragment Set Derivation}
	\label{sub-sec:break}
	
	After retrieving $\mathcal{R}_{cl}$ ($\forall cl\in\mathcal{T}$), corresponding valid (and more compact) edge fragment sets are obtained with two pruning techniques:
	1) {\em pairwise pruning} examines the pairs of edge fragments from consecutive $\mathcal{R}_{cl_i}$ and $\mathcal{R}_{cl_{i+1}}$ to eliminate the invalid edge fragments; 
	2) {\em sequence pruning} takes as input a sequence of consecutive edge fragment sets (after being filtered by pairwise pruning) to collectively eliminate the invalid edge fragment combinations. 

	To be precise, an edge fragment combination $EC=\langle ef_1, $ $\cdots, ef_{|EC|}\rangle$ is a time-ordered sequence of edge fragment, where for every pair of edge fragments $ef_i\in\mathcal{R}_{cl_i}$ and $ef_j\in\mathcal{R}_{cl_j}$ ($i<j$), $cl_i.t<cl_j.t$.
	An $EC$ is invalid if the minimum travel time between  any two adjacent $ef_i\in EC$ and $ef_{i+1}\in EC$ is greater than the corresponding time interval $\Delta t=cl_{i+1}.t-cl_i.t$. 
	Thus, an edge fragment $ef\in \mathcal{R}_{cl}$ is removed if it does not belong to any valid edge fragment combination.
	
\begin{figure}[!htb]
		\centering
		\vspace{-10pt}
		\subfigure[Pairwise Pruning]{
			\label{fig:tt-infer-l} 
			\includegraphics[width=0.45\columnwidth]{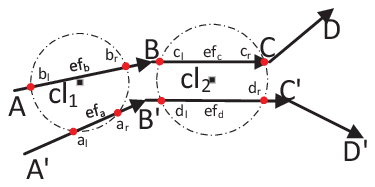}}
		\subfigure[Sequence Pruning]{
			\label{fig:tt-infer-r} 
			\includegraphics[width=0.45\columnwidth]{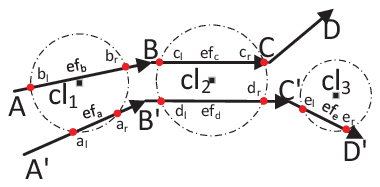}}
		\vspace{-10pt}
		\caption{Pruning Techniques}
		\vspace{-10pt}
		\label{fig:tt-infer}  
	\end{figure}

	Figure~\ref{fig:tt-infer-l} and~\ref{fig:tt-infer-r} illustrate the pairwise pruning and the sequence pruning, respectively.
	As shown in Figure~\ref{fig:tt-infer-l}, $cl_1$ and $cl_2$ are two consecutively observed cellular locations, 
	where $\mathcal{R}_{cl_1}=\{ ef_a, ef_b\}$, $\mathcal{R}_{cl_2}=\{ ef_c, ef_d\}$.
	With a globally specified maximum travel speed $v_{max}$, we determine that the traveling from $ef_{b}$ to $ef_{d}$ is invalid 
	and hence edge fragment combination $\langle ef_b, ef_d\rangle$ is invalid. 
	Similarly, $\langle ef_a, ef_c\rangle$ is invalid, leaving $EC_1=\langle ef_a, ef_d\rangle$ and $EC_2=\langle ef_b, ef_c\rangle$.
	Further, as shown in Figure~\ref{fig:tt-infer-r}, when $cl_3$ is considered, we have $\mathcal{R}_{cl_3}=\mathcal{R}_{cl_3}^\star=\{ef_e \}$.
	%
	%
	Through sequence pruning, we easily know that the only remaining valid edge combination is $EC_3=\langle ef_a, ef_d, ef_e \rangle$ which is extended by $EC_1$. 
	Moreover, as $EC_3$ is a most compact edge fragment set, we use it estimate the travel times w.r.t. $A^\prime B^\prime$, $B^\prime C^\prime$, and $C^\prime D^\prime$ (see Section~4.2).
	%
	
	
	\subsection{Object Motion Model}
	\label{sub:sec:state-duration-prob}

	As mentioned, at the heart of the trajectory cleansing process is a novel object motion model, namely, {\it Travel-time-aware Hidden semi-Markov Model} (TT-HsMM), which captures the movement of a mobile object on DTN. The idea is to consider the movement of the mobile object $o$ as the source of its cellular trajectory $o.\mathcal{T}$, observed/estimated within time window $\mathcal{W}$.
	Note that $o.\mathcal{T}$ contains a series of sub-trajectories, where each sub-trajectory is assumed to be generated from an edge $e$ of DTN, because $o$ cannot jump to another edge $e^\prime$ and jump back within a short time. 

	Accordingly, we model the movement of $o$ within $\mathcal{W}$ as a state-observation process.
	In this process, each state is an edge $e$ in DTN. Within the expected travel time of an edge $e_1$, a sub-trajectory $o.\mathcal{T}_1\subseteq o.\mathcal{T}$ is continuously observed from $e_1$, and then the sub-trajectory $o.\mathcal{T}_2\subseteq o.\mathcal{T}$ next to $o.\mathcal{T}_1$ is observed from another edge $e_2$ incident to $e_1$.
	The process of edge-subtrajectory interaction continues while $o.\mathcal{T}$ is observed.

\begin{figure}[!htb]
		\centering
		\vspace{-5pt}
		\subfigure[TT-HsMM]{
			\label{fig:tt-hsmm-top} 
			\includegraphics[width=0.3\columnwidth]{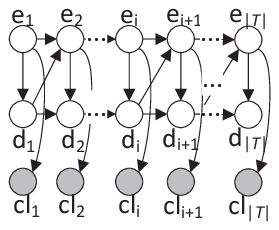}}
		\subfigure[Instantiation 1]{
			\label{fig:tt-hsmm-eg1} 
			\includegraphics[width=0.3\columnwidth]{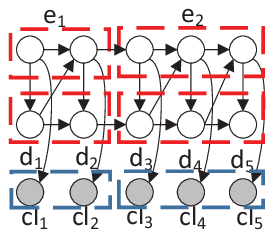}}
		\subfigure[Instantiation 2]{
			\label{fig:tt-hsmm-eg2} 
			\includegraphics[width=0.3\columnwidth]{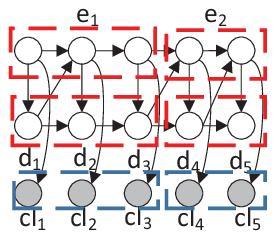}}
		\vspace{-5pt}
		\caption{Object Motion Model and Illustrations}
		\vspace{-10pt}
		\label{fig:hsmm-dbn} 
	\end{figure}

	Figure~\ref{fig:tt-hsmm-top} depicts the process of edge-subtrajectory interaction regarding a cellular trajectory $\mathcal{T}=\langle cl_1, cl_2, \cdots, cl_{|\mathcal{T}|}\rangle$, leveraging a Dynamic Bayesian Network. 

	As shown in Figure~\ref{fig:tt-hsmm-top}, three types of nodes are involved in this network:
	{\it edge nodes}, e.g., $e_i$, which denote the latent transportation edges; {\it remaining duration nodes}, e.g., $d_i$, which are the remaining traveling durations on corresponding edges; and {\it cellular location nodes}, e.g., $cl_i$, which are the observed cellular locations.
	Links between the nodes represent conditional dependencies.
	%
	
	Figure~\ref{fig:tt-hsmm-eg1} shows an instantiated TT-HsMM regarding a cellular trajectory $\mathcal{T}=\langle cl_1, cl_2, cl_3, cl_4, cl_5\rangle$.
	Here, the subtrajectory $\langle cl_1, cl_2 \rangle$ is associated with $e_1$ with the remaining durations $d_1$ and $d_2$, whilst $\langle cl_3, cl_4, cl_5 \rangle$ is assumed to be generated from $e_2$.
	Alternatively, Figure~\ref{fig:tt-hsmm-eg2} gives another instantiation. $\langle cl_1, cl_2\rangle$ is associated with $e_1$; the rest are associated with $e_2$.
	%
	To evaluate different instantiations, we need to compute the overall probability with respect to each instantiation, which reflects the object motion.
	
	
	The object motion over the DTN is described as follows. When a mobile object first enters a state (i.e., edge) $e$, the travel time $d$ is sampled from the travel time distribution $\mathcal{D}_e$ associated with $e$. 
	The state duration is determined by $d^\prime=d\times \frac{|ef.p_c,e.d|}{|e.s,e.d|}$, where $ef.eid=e.eid$, $ef\in\mathcal{R}_{cl_k}$ and $ef.p_c$ is $ef$'s center.
	Thus, $p(d_k=d^\prime|d_{k-1}=0,e_k=e)=p(t=d|\mathcal{D}_e)$, representing the probability that the travel time $d$ of $e$, is adopted.
	Thereafter, the remaining traveling duration deterministically subtracts the time gap between current cellular location and next cellular location, until the remaining traveling duration becomes less than $0$ after the subtraction. 
	During the subtraction process, when $d_{k-1}\geq \Delta_{k-1} t=cl_{k}.t-cl_{k-1}.t$, the state remains the same (i.e., the object still travels on the current edge).
	Accordingly, $e_{k-1}=e_k$, $p(e_k|e_{k-1},d_{k-1})=1$, and $p(d_k=d_{k-1}-\Delta_{k-1}t|d_{k-1},e_k)=1$.
	When $d_{k-1}$ is about to be negative, i.e., $d_{k-1}< \Delta_{k-1}t$, the state makes a stochastic transition to a new state (i.e., the mobile object should embark on a new edge $e_k=e^\prime$ incident to $e_{k-1}$ where $e^\prime.s=e_{k-1}.d$). 
	Thus, $p(e_k|e_{k-1},d_{k-1})= p(ef_n\in \mathcal{R}_{cl_k}|ef_m\in \mathcal{R}_{cl_{k-1}},\Delta_{k-1} t)$, where $ef_n.eid=e_k.eid$ and $ef_m.eid=e_{k-1}.eid$.
	After the transition, the travel time is again sampled from the embarked new edge $e_k$, the remaining traveling duration is re-computed. The process loops until it covers all the observed cellular locations.

	Based on TT-HsMM, we are able to compute two probabilities: $p(d_k=j|d_{k-1}=i, e_k=e)$ which corresponds to the links from one {\it duration node} $d_{k-1}$  to its adjacent {\it duration node} $d_k$ given the {\it edge node} $e_k$,  and $p(e_k=e_n|e_{k-1}=e_m, d_{k-1}=i)$ corresponding to the links from one {\it edge node} $e_{k-1}$ to its adjacent {\it edge node} $e_k$ given the {\it reaming duration node} $d_{k-1}$.
	
	\vspace{-15pt}
	\begin{equation*}
	\small
	\begin{multlined}
	p(d_k=j|d_{k-1}=i, e_k=e) = \\
	\begin{cases}
	p(t=d|\mathcal{D}_e)~~~\mathrm{if}\ i=0
	~~~(\mathrm{begin\ to\ travel\ on\ edge\ e})\\
	1~~~~~~~~~~~~~~~~~\mathrm{if}\ i\geq \Delta_{k-1}t=cl_k.t- cl_{k-1}.t
	~~~(\mathrm{traveling\ on\ }e)\\
	0~~~~~~~~~~~~~~~~~\mathrm{otherwise}
	\end{cases}
	\end{multlined}
	\end{equation*}
	\begin{equation*}
	\small
	\begin{multlined}
	p(e_k=e_n|e_{k-1}=e_m, d_{k-1}=i) = \\
	\begin{cases}
	1~~~~~~~~~~~~~~~~~~~~\mathrm{if}\ i\geq \Delta_{k-1}t\ ~~~~~~(\mathrm{object\  still\ travels\ on\ }e_m)\\
	p(ef_n\in \mathcal{R}_{cl_k}|ef_m\in \mathcal{R}_{cl_{k-1}},\Delta_{k-1} t) \\
	~~~~~~~~~~~~~~~~~~~~~\mathrm{if}\ i<\Delta_{k-1} t 
	~~(\mathrm{transition\ is\ required})
	\end{cases}
	\end{multlined}
	\end{equation*}
	
	Note that, the time gaps between two consecutive cellular locations may be varied due to the missing locations.
	Additionally, the required probabilities can be computed in terms of $p(t=d|\mathcal{D}_e)$ which can be derived from the learned travel time distribution $\mathcal{D}_e$ (cf. Section~\ref{subsec:de}), and the adaptive transition probability $p(ef_n\in \mathcal{R}_{cl_k}|ef_m\in \mathcal{R}_{cl_{k-1}},\Delta_{k-1} t)$ which can be approximated (cf. Section~\ref{sub-sec:hsmm}).

	Another probability required in the instantiation evaluation is $p(cl|e)$.
	Since an edge fragment uniquely corresponds to one transportation edge and the physical location can only fall into the derived edge fragment, $p(cl|e)=p(cl|ef)$ when $e.eid=ef.eid$.
	And $p(cl|ef)$ is the {\it robust emission probability}, which can be directly computed (cf. Section~\ref{sec:robust-adaptive}).
	Therefore, after the valid edge fragment sets corresponding to the cellular trajectory of $\mathcal{T}=\langle cl_1,\cdots,cl_{|\mathcal{T}|}\rangle$ are derived, all the probabilities required in its TT-HsMM can be computed.
	%
		
	\subsection{Traveled Edge Sequence Inference}
		
	Based on the TT-HsMM, the edge sequence inference aims to find the {\em most probable edge sequence} the object travelled on.\footnote{The edge sequence is also referred to as a {\em path} in the paper and used interchangeably.} 
	The most probable edge sequence $e_{1:|\mathcal{T}|}$, corresponding to a cellular trajectory $\mathcal{T}=\langle cl_1,\cdots,cl_{|\mathcal{T}|}\rangle$,  is obtained by maximizing the joint probability $p(cl_{1:|\mathcal{T}|},$ $e_{1:|\mathcal{T}|})$. 
	Here, $cl_{1:|\mathcal{T}|}$ denotes the cellular location sequence in $\mathcal{T}$. 
	\begin{equation}
	e_{1:|\mathcal{T}|}=\argmax_{e_{1:|\mathcal{T}|}} p(cl_{1:|\mathcal{T}|}, e_{1:|\mathcal{T}|})
	\end{equation}
	%
	\begin{theorem}
		\label{th:most-prob}
		The optimization problem can be solved by dynamic programming.
	\end{theorem}
	\textit{Proof:  } Guided by the Bayesian Network in Figure~\ref{fig:hsmm-dbn}, the probability $p(cl_{1:|\mathcal{T}|}, e_{1:|\mathcal{T}|})$ can be written as follows.
	\begin{equation*}
	\begin{multlined}
	p(cl_{1:|\mathcal{T}|}, e_{1:|\mathcal{T}|})\\
	\hspace{-25pt}=\prod_{i=2}^{|\mathcal{T}|}p(cl_i|e_i)\times \prod_{i=2}^{|\mathcal{T}|} p(d_i|d_{i-1},e_i)\times
	\prod_{i=2}^{|\mathcal{T}|} p(e_i|e_{i-1},d_{i-1})\\
	\hspace{-100pt}\times p(cl_1|e_1)\times p(d_1|e_1)\times p(e_1)\\
	=\prod_{i=2}^{|\mathcal{T}|-1}p(cl_i|e_i)\times \prod_{i=2}^{|\mathcal{T}|-1} p(d_i|d_{i-1},e_i)\times
	\prod_{i=2}^{|\mathcal{T}|-1} p(e_i|e_{i-1},d_{i-1})\\
	\hspace{-40pt}
	\times p(cl_1|e_1)\times p(d_1|e_1)\times p(e_1)\times p(cl_{|\mathcal{T}|}|e_{|\mathcal{T}|}) \\ 
	\hspace{-40pt} \times p(d_{|\mathcal{T}|}|d_{|\mathcal{T}|-1},e_{|\mathcal{T}|})\times 
	p(e_{|\mathcal{T}|}|e_{|\mathcal{T}|-1},d_{|\mathcal{T}|-1})\\
	=p(cl_{1:|\mathcal{T}-1|},e_{1:|\mathcal{T}-1|})\times
	p(cl_{|\mathcal{T}|}|e_{|\mathcal{T}|})\times p(d_{|\mathcal{T}|}|d_{|\mathcal{T}|-1},e_{|\mathcal{T}|})\\ \hspace{-120pt}\times p(e_{|\mathcal{T}|}|e_{|\mathcal{T}|-1},d_{|\mathcal{T}|-1})\\
	\end{multlined}
	\end{equation*}
	
	As shown, the joint probability can be recursively expressed in terms of the edge-based robust emission probability $p(cl_{i}|e_i)$ (where $p(cl_{i}|e_i)=p(cl_i|ef_i)$) and the two conditional probabilities $p(d_i|d_{i-1},e_i)$ and $p(e_i|e_{i-1},$ $d_{i-1})$ regarding adaptive transitions.
	As previously discussed, all these probabilities can be computed directly.$~~~~~~~~~~~~~~~~~~~~~~~~~~~~~~~~~~~\square$
	
	Therefore, similar to the Viterbi algorithm used in regular HMM, we develop an algorithm for TT-HsMM to infer the most probable edge sequence. 	Algorithm~\ref{alg:viterbi-hsmm-mv} shows the pseudocode.
	%
	%
	Initially (lines 1-3), we assign the computed robust emission probability to each edge in $\mathcal{R}_{cl_1}$,
	and the back pointers zeros. 
	The recursion step (lines 4-7) applies the equation in Theorem~\ref{th:most-prob} to incrementally solve the subproblems.
	Finally (lines 8-10), the most probable edge sequence is found according to the  back pointers.
	
	\begin{algorithm}[!htp]
		\LinesNumbered
		\SetKwInOut{Input}{Input}
		\SetKwInOut{Output}{Output}
		\SetKw{Return}{return}
		\caption{{\bf Most Probable Edge Sequence Inference}}
		\label{alg:viterbi-hsmm-mv}
			\tcc{initialization step}
		\For{each edge $e_1 \supseteq ef_1\in \mathcal{R}_{cl_1}$}
		{
			\textit{viterbi[$e_1$,1]}$\leftarrow  p(cl_1|ef_1)$;\\
			\textit{backpointer[$e_1$,1]}$\leftarrow 0$;\\
		}
		\tcc{recursion step}
		\For{each time step $k$ from $2$ to $|\mathcal{T}|$}
		{
			\For{each edge $e_k\supseteq ef_k\in \mathcal{R}_{cl_k}$}
			{
				\textit{viterbi[$e_k$,k]}$\leftarrow  \mathrm{max}_{d_k\sim \mathcal{D}_{e_k}}$\textit{viterbi}$[e_{k-1},k-1]*p(e_k|e_{k-1}, d_{k-1})*p(d_k|d_{k-1},e_k)* p(cl_{k}|e_k)$;\\
				\textit{backpointer[$e_k$,$k$]}$\leftarrow \argmax_{d_k\sim \mathcal{D}_{e_k}}$\textit{viterbi}$[e_k,t-d]*p(e_k|e_{t-d}, d_{t-d})*p(d|d_{t-d},e_k)$;\\
			}
		}
		\tcc{termination step}
		\textit{viterbi}$[e_{|\mathcal{T}|+1},|\mathcal{T}|+1]\leftarrow \max_{e\supseteq ef\in \mathcal{R}_{cl_{|\mathcal{T}|}}}$\textit{viterbi}[$e_{|\mathcal{T}|}$,$|\mathcal{T}|$];\\
		\textit{backpointer}$[e_{|\mathcal{T}|+1},|\mathcal{T}|+1]\leftarrow\argmax_{e\supseteq ef\in \mathcal{R}_{cl_{|\mathcal{T}|}}}$\textit{viterbi}[$e_{|\mathcal{T}|}$,$|\mathcal{T}|$];\\
		\Return{the most probable edge sequence by following backpointers from \textit{backpointer}$[e_{|\mathcal{T}|+1},|\mathcal{T}|+1]$};
	\end{algorithm}
	
	It takes $O(|\mathcal{R}_{cl_i}|\times |\mathcal{D}_{e_i}|)$ time to compute the forward variable for a single iteration, where the cardinality of $\mathcal{R}_{cl_i}$ is the number of contained edge fragments and the cardinality of $\mathcal{D}_{e_i}$ is the number of outcomes contained in $\mathcal{D}_{e_i}$.
	Hence, the time complexity is $O(|\mathcal{R}_{cl_i}|_{max} \times |\mathcal{D}_{e}|_{max} \times |\mathcal{T}|)$, where $ |\mathcal{R}_{cl_i}|_{max}$ denotes the largest cardinality of $\mathcal{R}_{cl_i}$ for any cellular location in $\mathcal{T}$, $|\mathcal{D}_{e}|_{max}$ denotes the largest cardinality of $\mathcal{D}_{e}$ for any edge included, and $|\mathcal{T}|$ is the number of cellular locations in the trajectory $\mathcal{T}$.
	
	As $|\mathcal{D}_{e}|$ and $|\mathcal{R}_{cl_i}|$ are two factors determining the time complexity, we pursue compact $\mathcal{D}_e$ and compact $\mathcal{R}_{cl}$, as described in Section~\ref{subsec:de} and~\ref{sub-sec:break}, respectively.

	\subsection{Physical Location Inference}
	\label{sub:sec:physical-location}
	
	After obtaining the most probable edge sequence $\mathcal{S}=\langle e_1, e_2, \cdots, e_{|\mathcal{T}|} \rangle$ for a cellular trajectory $\mathcal{T}=\langle cl_1,$ $cl_2, \cdots, cl_{|\mathcal{T}|}\rangle$, we can track back to the corresponding edge fragment sequence $\mathcal{S}^\prime=\langle ef_1, ef_2, \cdots, ef_{|\mathcal{T}|} \rangle$, where $ef_i.eid=e_i.eid$.
	
	For a cellular locations $cl_i\in \mathcal{T}$, there exists a unique $ef_i$ correspondingly.
	In this case, the center of the edge fragment approximates its physical location.
	We replace $cl_i$ by  $ef.p_c=\frac{ef_i.p_l+ef_i.p_r}{2}$.
	For the missing values between two cellular locations $cl_i$ and $cl_{i+1}$, we track back to the corresponding particles, denoted by $\mathcal{B}$, that successfully simulate the movement between $ef_i\in\mathcal{S}$ and $ef_{i+1}\in\mathcal{S}$.
	A particle $par\in\mathcal{B}$ is randomly selected.
	After that, the missing location at time $t$ ($cl_i.t<t<cl_{i+1}.t$) is filled with the particle $par$'s location at the corresponding moment.
		
	\vspace{-3pt}
	\section{Empirical Evaluation}
	\label{sec:evaluation}
	\vspace{-3pt}
	
	\subsection{Experiment Setup}
	
	\textbf{Transportation network:} We obtain the transportation network of Singapore 
	from OpenStreetMap\footnote{https://www.openstreetmap.org/}.
	The network, containing $285,102$ vertices and $342,261$ edges, has different edge types, including motorway, trunk, subway, etc.
	%
	Moreover, a grid index with cell length $100$~meters is built to support efficient range queries. 
	
	\textbf{Cellular Trajectories with ``Ground Truth''}: To evaluate the accuracy of the inferred physical locations, we collect GPS locations by smartphones along with the cellular locations 
	from StarHub, a major cellular network operator in Singapore and South-east Asia. We align the GPS locations to network edges using HMM~\cite{newson2009hidden} and use them  
	as the ground truth for evaluation.
	Ten cellular trajectories $\mathcal{T}_1-\mathcal{T}_{10}$ are collected and named asdataset $D_1$.	The numbers of cellular locations in $D_1$ are summarized in Table~\ref{table:trajs}.

	\begin{table}[!htb]
		\centering
		\scriptsize
		\begin{tabular}{|c|c|c|c|c|c|c|c|c|c|c|}
			\hline
			$\mathcal{T}_1$ & $\mathcal{T}_2$ & $\mathcal{T}_3$ & $\mathcal{T}_4$  & $\mathcal{T}_5$ & $\mathcal{T}_6$ & $\mathcal{T}_7$ & $\mathcal{T}_8$ & $\mathcal{T}_{9}$ & $\mathcal{T}_{10}$ \\
			\hline
			$1552$ & $1913$ & $570$ & $304$ & $1862$ & $286$ & $669$ & $717$ & $790$ & $2400$ \\
			\hline
		\end{tabular}
		\vspace{-6pt}
		\caption{Number of Cellular locations in $D_1$}\label{table:trajs}
		\vspace{-10pt}
	\end{table}

	%
	The average spatial uncertainty ($u$) in $D_1$ is $3.82$, and proportion of the cellular locations with $u=4$ and $u=5$ is over $64\%$.
	%
	%
	Meanwhile, the average time gap between two consecutively observed cellular locations is $13.82$ seconds.
		
	\textbf{Cellular Trajectories without Ground Truth}: To generate the dynamic transportation network, and evaluate the efficiency of the proposals, we obtain 
	cellular locations observed within the same times and same regions of $D_1$ from StarHub.
	%
	This dataset, denoted by $D_2$, contains $1,790,042$ cellular locations observed from $2,031$ mobile objects.
	%
	%
	On average, the uncertainty degree is $3.93$ and the time gap between two consecutive cellular locations is $14.5$~seconds.

	\textbf{Baselines:} 
	As discussed in Section~\ref{sec:related}, 
	we extend six map matching or path recovery methods to support {\it CTC}, including STRS~\cite{DBLP:conf/kdd/WuMSZZCW16}, SnapNet~\cite{mohamed2016accurate}, HMM~\cite{raymond2012map}, PF~\cite{kempinska2016probabilistic}, OHMM~\cite{goh2012online}, and CTrack~\cite{thiagarajan2011accurate}.

	In order to produce the locations,
	%
	interpolations on the derived edge sequence are used for the missing locations.
	For CTrack~\cite{thiagarajan2011accurate} and OHMM~\cite{goh2012online}, the required real-time speed, acceleration, and heading direction are calculated from the raw cellular trajectories. 
	
	\textbf{Implementation Details:}
	All algorithms are implemented in JAVA $1.8.0\_45$ under Linux Ubuntu 16.04.
	All experiments are conducted on a server with 16 GB main memory and 3.6 GHz E3-1271 v3 Intel(R) Xeon(R) CPU.
	
	\vspace{-4pt}
	\subsection{Overall Evaluation}
	\vspace{-4pt}
	\subsubsection{Effectiveness}
	
	We conduct experiments on $D_1$ to evaluate the effectiveness of DTNC.
	We use the {\em Euclidean deviation} $d=dist(cl^\prime, gps)$ between an inferred physical location $cl^\prime$
	and its corresponding GPS location $gps$ with the same timestamp to measure the {\em spatial error}.

	\begin{figure}[!htb]
		\centering
		\vspace{-5pt}
		\includegraphics[width=0.49\textwidth]{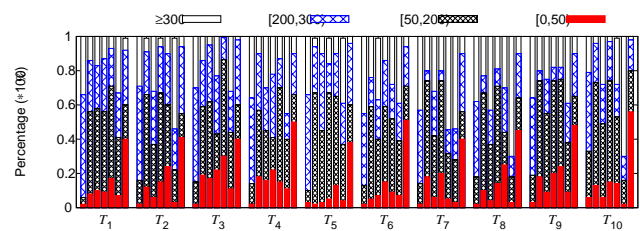}
		\vspace{-15pt}
		\caption{Overall Results}	\label{fig:eff-acc-varying-data}
		\vspace{-10pt}
	\end{figure}
	
	Figure~\ref{fig:eff-acc-varying-data} shows the results 
	grouped by the trajectories. Each group has seven stacked histograms, corresponding to the methods under evaluation, including (from left to right) PF, HMM, OHMM, STRS, SnapNet,  CTrack, and DTNC.
	%
	%
	As shown, DTNC produces the best cleansing results for all the trajectories (from $\mathcal{T}_1$ to $\mathcal{T}_{10}$).
	In particular, after cleansing by DTNC, each cellular trajectory has more than $40\%$ cellular locations with spatial error within $50$~meters, significantly better than other methods.
	On average, DTNC has less than $10\%$ cleansed cellular locations with spatial errors greater than $300$~meters (i.e., the locations with uncertainty degree $4$ or $5$), dropped from $64\%$ in the original dataset. 
	However, the other methods have over $25\%$ cleansed cellular locations with spatial errors in this range.
	%
	
	\begin{figure*}[!htbp]
		\centering
		\vspace{-5pt}
		\includegraphics[width=1.9\columnwidth]{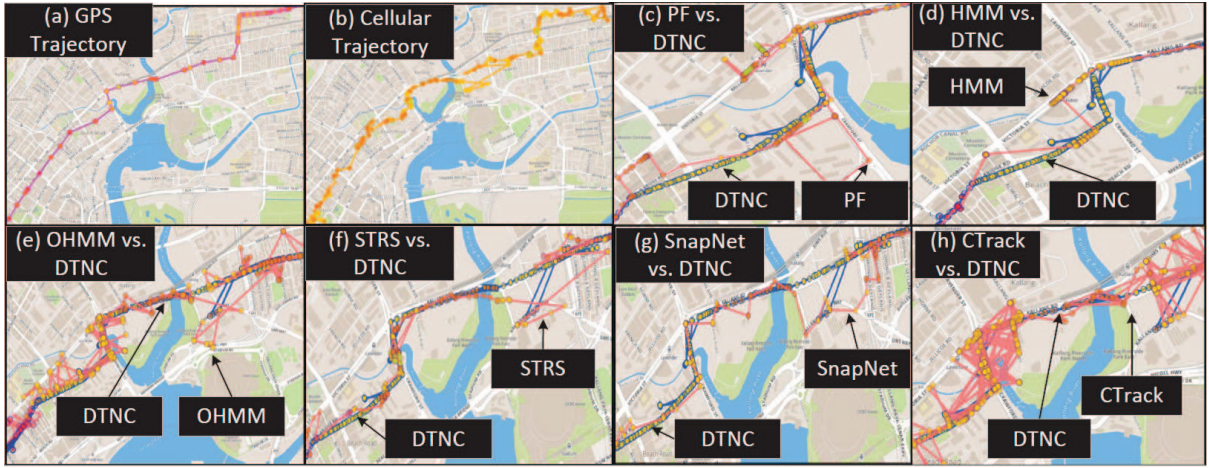}
		\caption{Detailed Comparison}	
		\vspace{-15pt}
		\label{fig:eff-detail}
	\end{figure*}
	
	The reasons that DTNC outperforms these baselines are two-fold.
	First, DTNC more accurately infers the traveled edge sequence. 
	It uses robust emission probabilities to avoid overemphasis on the Euclidean distances between a cellular location and its surrounding edges; uses adaptive transition probabilities to capture various travel times on edges and avoid unnecessary detours; and uses TT-HsMM, which considers the duration for each edge, to infer the state sequence. 
	Second, DTNC more accurately estimates the physical locations. 
	By considering the real-time travel times of edges, it infers the missing locations based on the traffic-aware simulations.

	\textbf{Case Study}.
	We present a case study in Figure~\ref{fig:eff-detail} to illustrate how and why DTNC excels in \textit{CTC} and significantly outperforms its potential rivals.
	Figure~\ref{fig:eff-detail}(a)-(b) shows a GPS trajectory (ground truth) and the corresponding raw cellular trajectory, respectivlely. Figure~\ref{fig:eff-detail}(c)-(h) show head-to-head comparisons between DTNC and other methods. 
	
	PF~\cite{kempinska2016probabilistic} estimates cellular locations by Monte Carlo simulations that dispatch particles according to the spatial coherence. However, owing to the high spatial error of cellular locations, the particles may be led to wrong directions and thus inaccurate cellular locations. As shown,  PF easily gets affected by the noisy cellular locations (see in Figure~\ref{fig:eff-detail}(c)). 
	
	HMM~\cite{newson2009hidden} considers the spatial aspect of the trajectory, trying to align the cellular trajectory with the underlying network.
	As shown, HMM generates some unnecessary detours because the high spatial errors drive the inferred routes towards wrong edges (see in Figure~\ref{fig:eff-detail}(d)).
	
	OHMM~\cite{goh2012online} relies on the real-time sensor data, i.e., speed and direction, on smartphones to compute the speeding penalty factor, momentum change, and the sensor-deduced traveling distance to calculate its distance discrepancy. For cellular data, such information can only be derived from the raw cellular trajectory, resulting in inaccuracy. As shown, OHMM produces chaotic locations when the required real-time speed and heading direction are computed from the raw cellular trajectory (see in Figure~\ref{fig:eff-detail}(e)).
	
	STRS~\cite{DBLP:conf/kdd/WuMSZZCW16} relies on the HMM~\cite{newson2009hidden} results to train a regression model for the travel time estimation, and recovers a route. 
	However, the HMM results are erroneous for cellular trajectories, leading to the faulty routes and locations.
	As shown, since STRS uses HMM-aligned cellular trajectories to train its travel time regression model, it gives fuzzy paths similar to its training trajectory (see in Figure~\ref{fig:eff-detail}(f)).	

	SnapNet~\cite{mohamed2016accurate} applies customized speed filter, $\alpha$-trimmed mean filter, and direction filter.
	However, in presence of long high spatial error in CTC, and lacks of road network exploration, these filters sometimes rule out actual edges, rendering wrong inferred locations.
	As shown, SnapNet's filters rule out correct edges when the high-spatial-error locations (whose uncertainty degrees are greater than four) appear consecutively (see in Figure~\ref{fig:eff-detail}(g)). 

	CTrack~\cite{thiagarajan2011accurate} leverages the cellular fingerprints collected from smartphones, and the ground truth training data to infer the location grids generating these fingerprints. 
	For cellular data, such information can only be approximately derived, leading to inaccurate results.
	As shown, CTrack produces very messy locations when we replace its GPS fingerprints by the cellular locations (in Figure~\ref{fig:eff-detail}(h)).

	In summary, DTNC convincingly dominates these state-of-the-art methods for \textit{CTC}.	
	
	\begin{figure*}[!htbp]
		\begin{minipage}[b]{0.49\columnwidth}
			\subfigure{
				\includegraphics[width=\columnwidth,height=3.3cm]{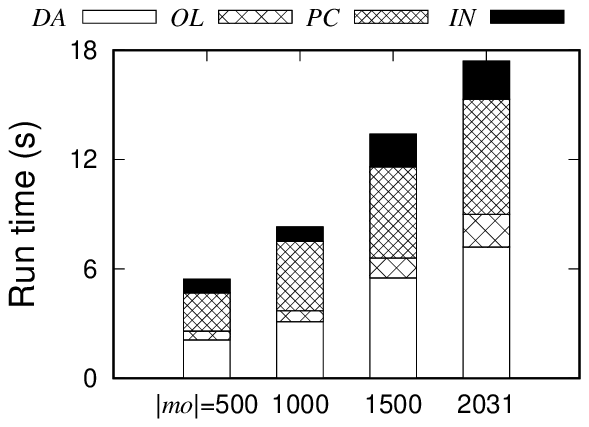}}
			\vspace{-8pt}
			\caption{Runtime Analysis}
			\label{fig:rt-ana}
		\end{minipage}
		\begin{minipage}[b]{0.49\columnwidth}
			\centering
			\includegraphics[width=\columnwidth,height=3.3cm]{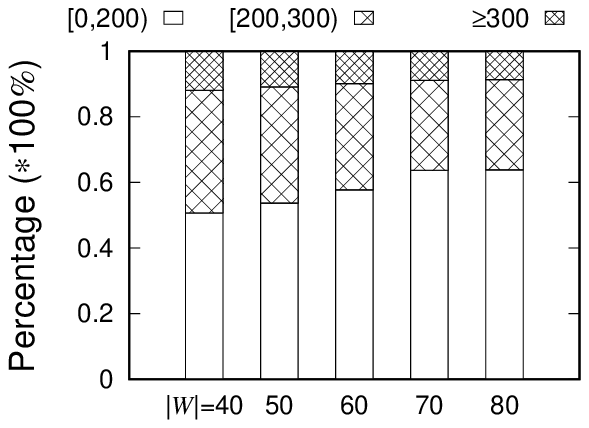}
			\vspace{-8pt}
			\caption{Varying $|\mathcal{W}|$}
			\label{fig:varying-w}
		\end{minipage}
		\hfill
		\begin{minipage}[b]{0.49\columnwidth}
			\centering
			\includegraphics[width=\columnwidth,height=3.3cm]{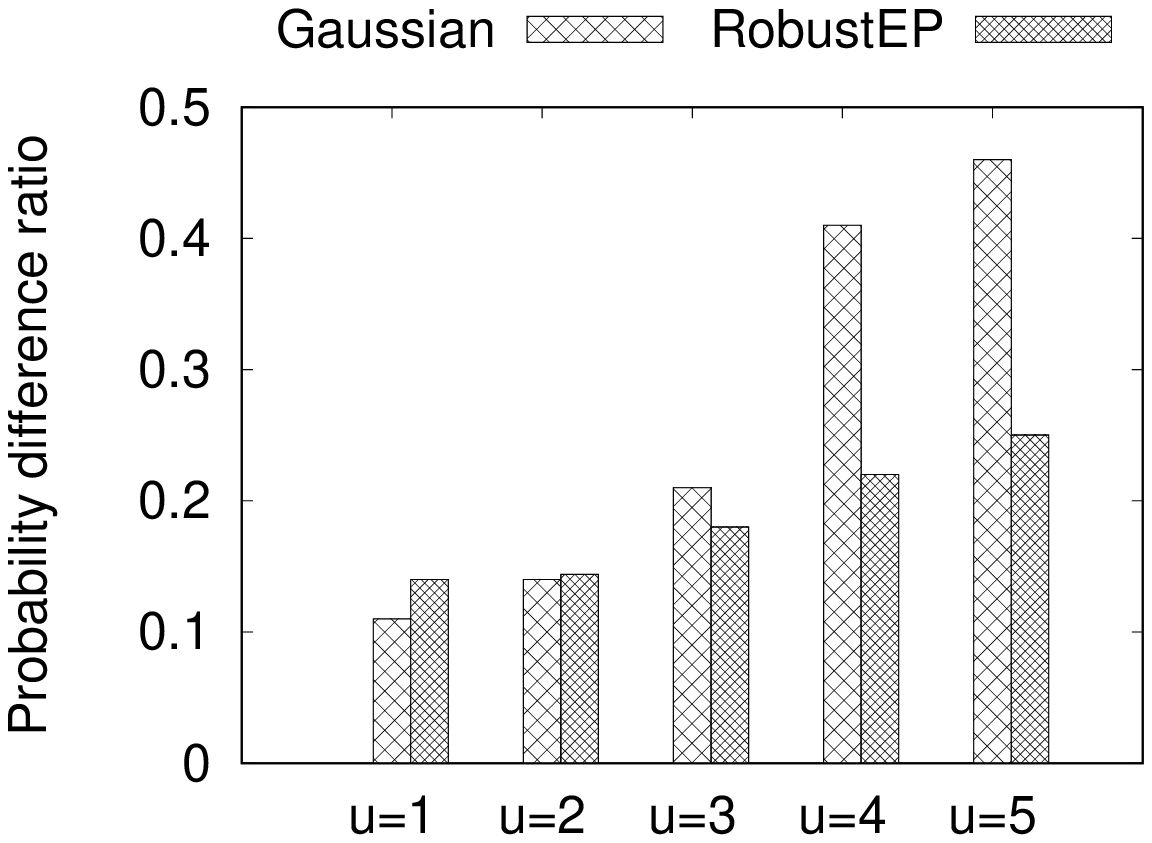}
			\vspace{-8pt}
			\caption{$PDR^{EP}$}
			\label{fig:eva-emis:vary-u}
		\end{minipage}
		\begin{minipage}[b]{0.49\columnwidth}
			\centering
			\includegraphics[width=\columnwidth,height=3.3cm]{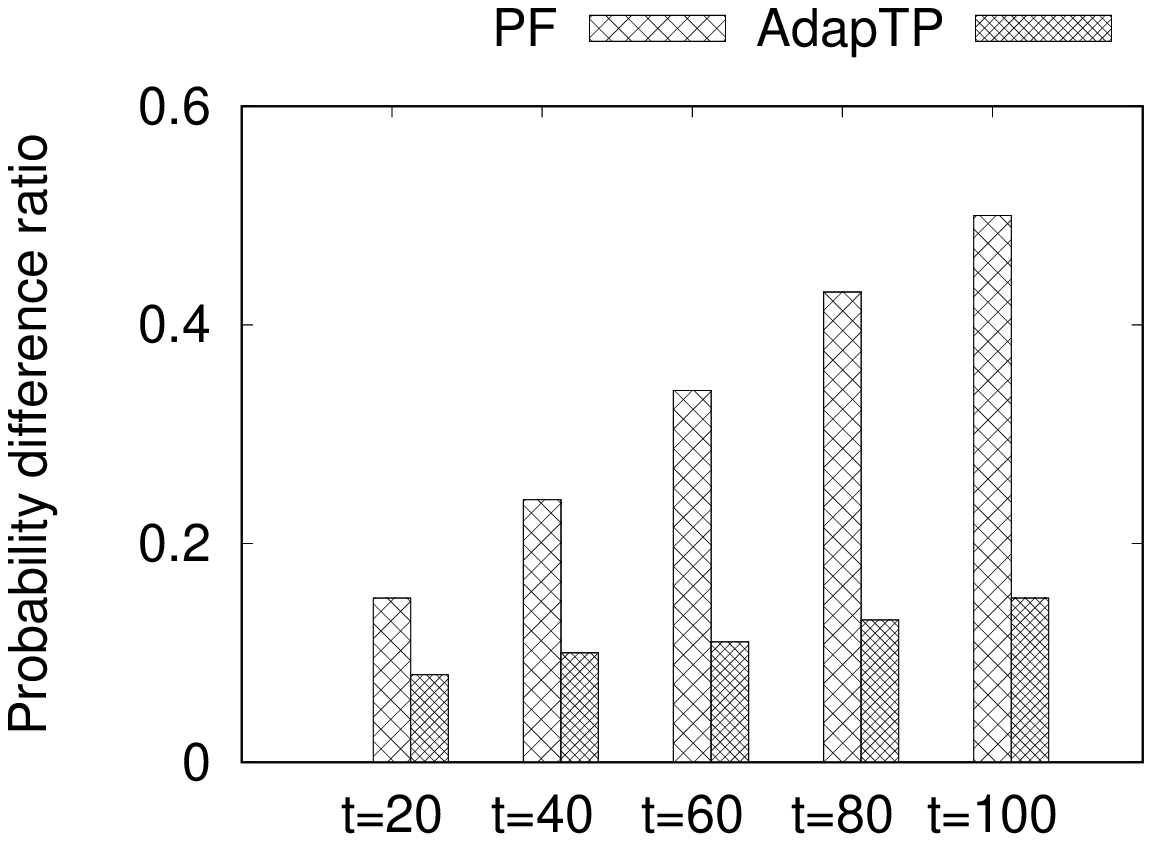}
			\vspace{-8pt}
			\caption{$PDR^{TP}$}
			\label{fig:trans-prob:vt}
		\end{minipage}
		
	\end{figure*}

	\begin{figure*}[!htbp]
		\begin{minipage}[b]{0.5\textwidth}
			\centering
			\subfigure[Pairwise Pruning]{
				\label{fig:two-f:pf}
				\includegraphics[width=0.48\textwidth,height=3.3cm]{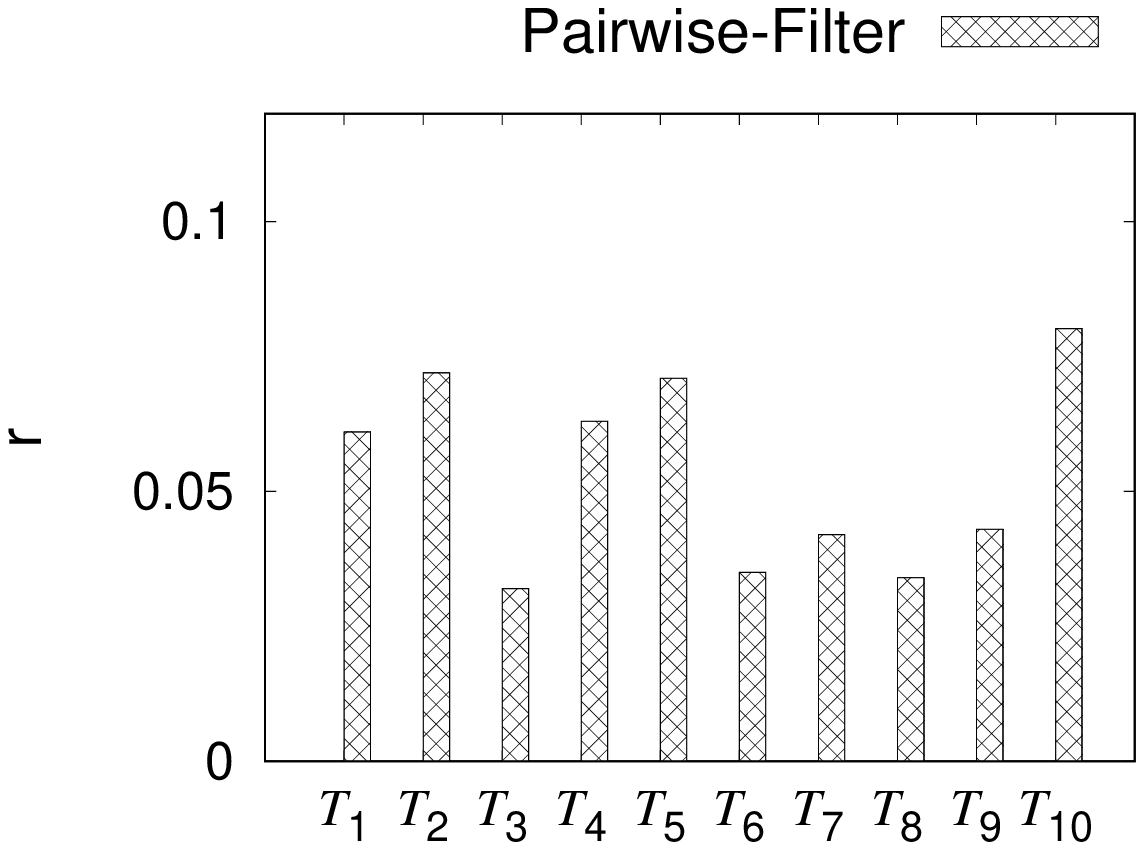}}
			\subfigure[Sequence Pruning]{
				\label{fig:two-f:sf}
				\includegraphics[width=0.48\textwidth,height=3.3cm]{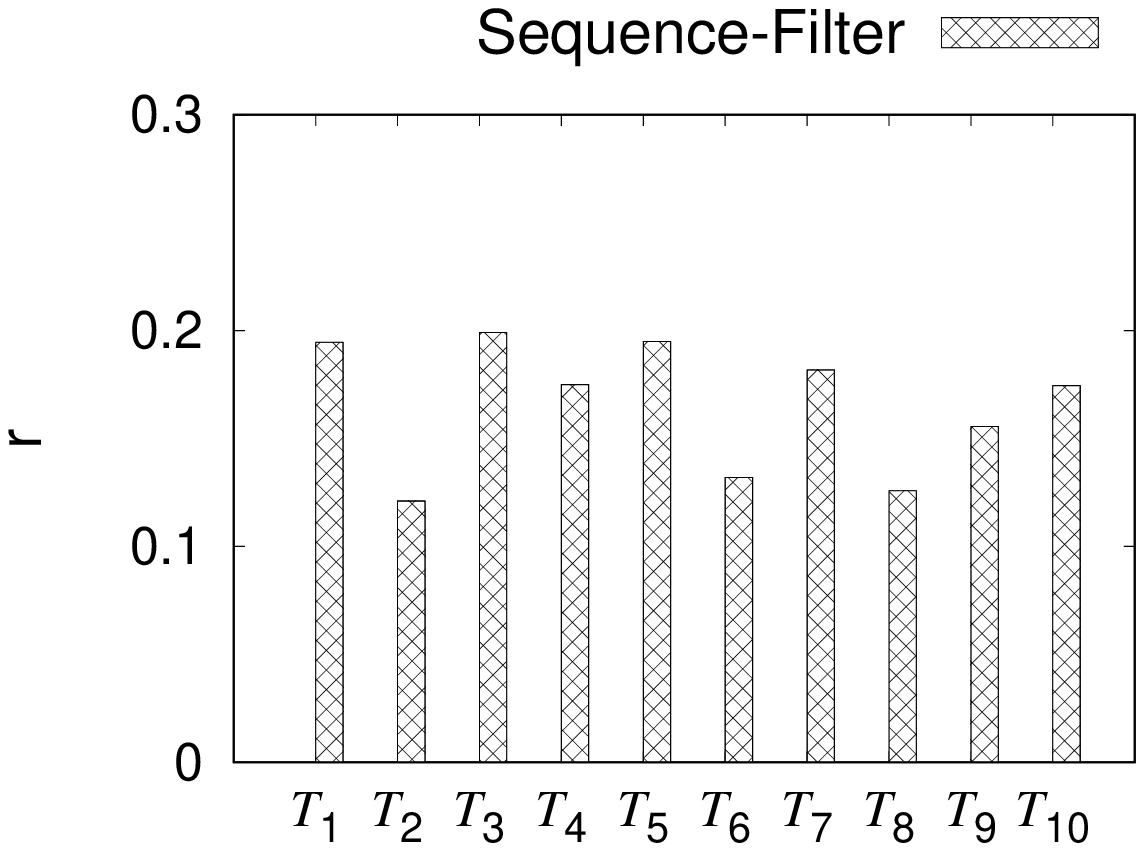}}
			\vspace{-10pt}
			\caption{Two Filters}
			\label{fig:two-f}
		\end{minipage}
		\hfill
		\begin{minipage}[b]{0.5\textwidth}
			\centering
			\subfigure[Varying $\delta$]{
				\label{fig:de-learn:ti}
				\includegraphics[width=0.48\textwidth,height=3.3cm]{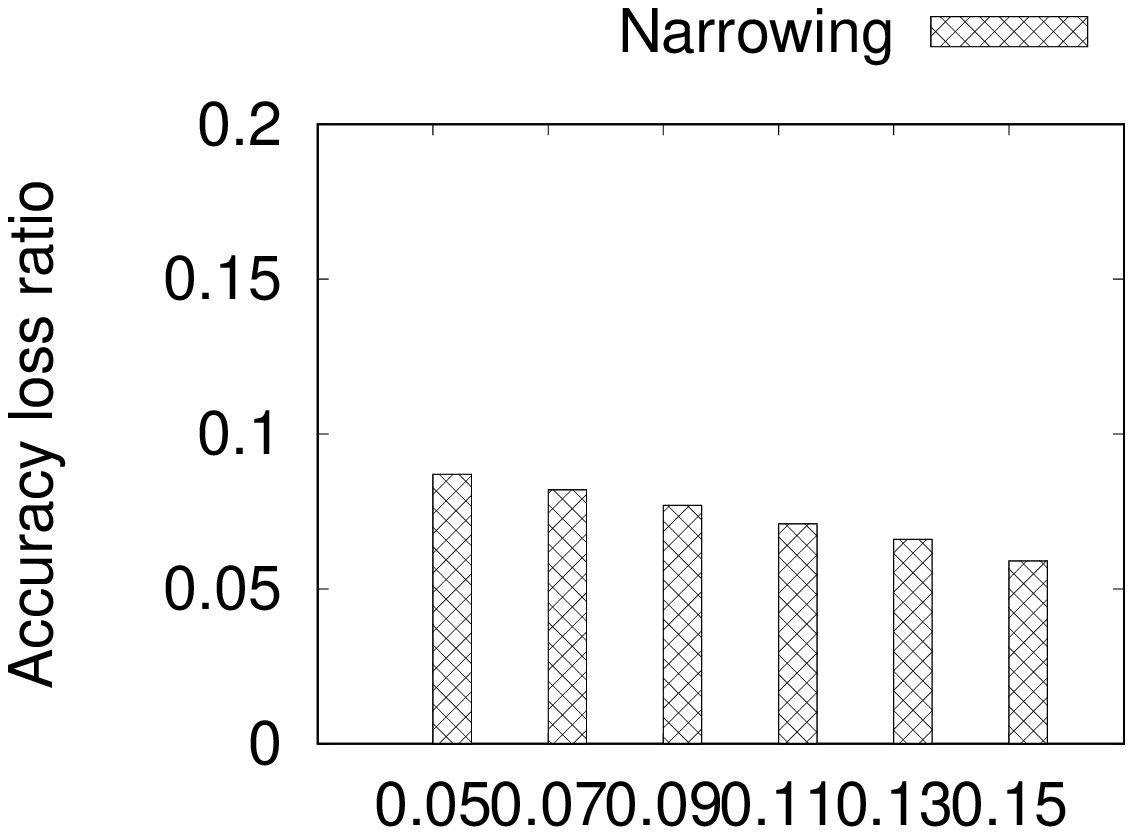}}
			\subfigure[Varying $\epsilon$]{
				\label{fig:de-learn:varyd}
				\includegraphics[width=0.48\textwidth,height=3.3cm]{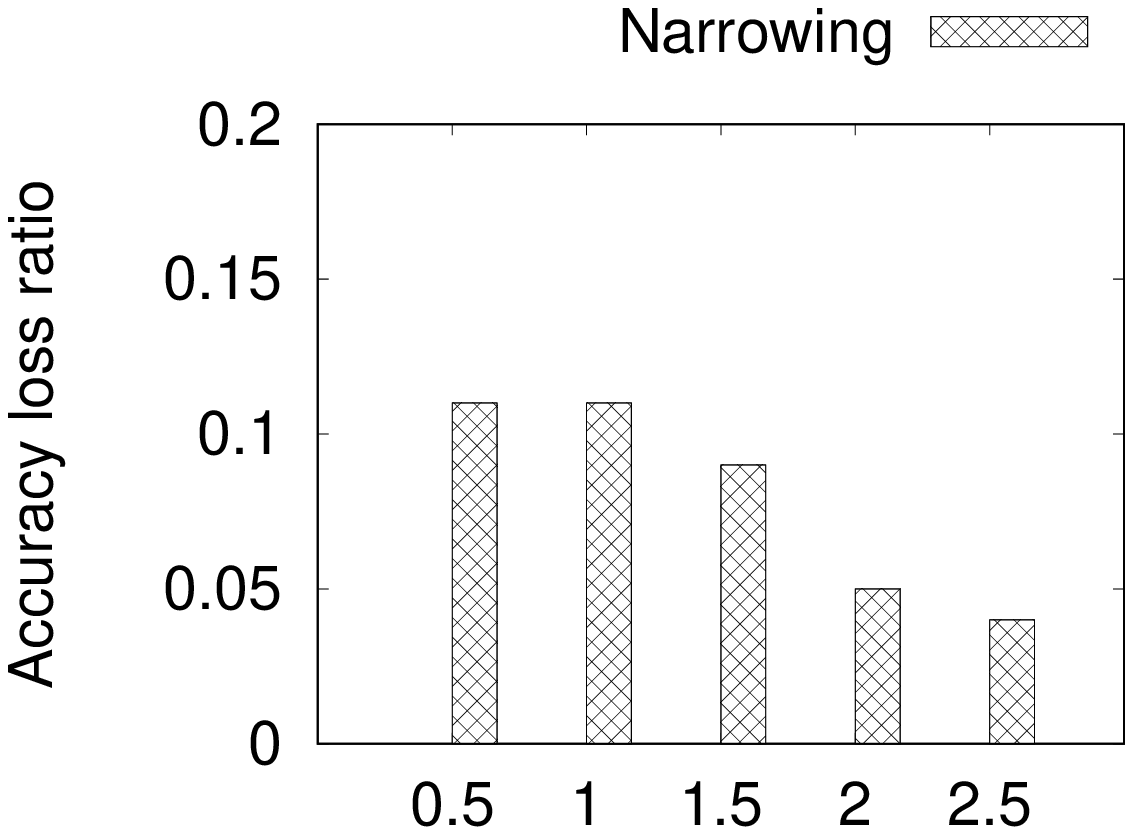}}
			\vspace{-10pt}
			\caption{$\mathcal{D}_e$ Learning}
			\label{fig:de-learn}
		\end{minipage}
	\end{figure*}

	\subsubsection{Efficiency}
	
	We also measure the run times to evaluate the efficiency of DTNC using $D_2$.
	Since the cleansing can be parallelized in terms of the cellular trajectories, $8$ threads are used to cleanse the trajectories from $2031$ mobile objects collected within one hour. 
	
	Figure~\ref{fig:rt-ana} reports the detailed DTNC running time which consists of four parts:
	\textit{DA} denotes valid edge fragment computation, including $\mathcal{R}_{cl}$ retrieval, and pairwise pruning and sequence pruning;
	\textit{OL} denotes online $\mathcal{D}_e$ learning and update;
	\textit{PC} represents probability computation, including robust emission probability computing and transition probability approximation; and \textit{IN} represents HsMM-MV construction, edge sequence inference, and location inference.
	As shown, when the number of mobile objects (denoted by $|mo|$) increases, the run time grows.
	\textit{DA} takes the largest portion of run time due to the range query for each cellular location
	and the further pruning.
	\textit{PC} takes the second largest portion of run time, because it not only computes emission probability but also performs Monte Carlo simulations to approximate a transition probability.
	%
	%
	Nevertheless, thanks to dynamic transportation network and the Dirichlet prior, \textit{PC} itself is much more efficient than the conventional particle filter.
	Run time of \textit{IN} consists of (i) the optimal sequence inference, and (ii) the physical location estimation.
	For a long cellular trajectory, \textit{IN} often only need to infer the most probable edge sequence for small sub-trajectories due to existence of the most compact $\mathcal{R}^\star$ sequences (occupying around $15\%$ of the cellular locations).
	Therefore, \textit{IN} is scalable w.r.t. the number of trajectories (and mobile objects).
	Due to the well-devised narrowing method, a small travel time range is maintained for each $\mathcal{D}_e$. Hence, \textit{OL} is scalable w.r.t. the number of mobile objects.
	
	\vspace{-5pt}
	\subsection{Evaluation of DTNC Components}
	
	In this section, we evaluate the individual components in DTNC and carry out a sensitivity test on window size. 
	
	\textbf{Robust Emission Probability}: We show that the devised robust emission probability normally does not assign a higher probability value to a transportation edge not been actually traveled. 
	We compare it with the widely used Gaussian distribution $\frac{1}{\sqrt{2\pi}\delta}e^{-0.5(\frac{dist(cl_t,X_t)}{\delta})^2}$.
	Leveraging $D_1$, the accuracy of the emission probability can be evaluated by the {\em probability difference ratio} for emission probability ($PDR^{EP}$): $\frac{p_{max}(cl|e)-p_{a}(cl|e)}{p_{max}(cl|e)}$, where $p_{a}(cl|e)$ is the emission probability of the actually traveled edge $e$, and $p_{max}(cl|e)$ is the maximum emission probability value, for all edges in $\mathcal{R}_{cl}$.
	A smaller probability difference ratio for emission probability is preferred.
	Figure~\ref{fig:eva-emis:vary-u} shows the average  probability difference ratios by varying uncertainty degrees (from $u=1$ to $u=5$) for the cellular locations in $D_1$.
	We can see that the proposed robust emission probability (labeled by RobustEP) achieves much lower probability difference ratios, compared with the Gaussian distribution with $50$~meters variance. This suggests that RobustEP tends to assign a low probability to an edge which is not actually traveled.
	Further, as $u$ increases, the probability difference ratios of RobustEP increase gently. In contrast, the Gaussian distribution is easily influenced by a high $u$.

	\textbf{Adaptive Transition Probability}: We show that the adaptive transition probability (labeled by AdapTP) is able to describe the long-distance movement by assigning a higher probability to the actually traveled path.
	We compare it with the particle filter, which does not consider the real-time traffic conditions and depends on the observed cellular locations to update the weights of particles.
	Leveraging $D_1$,  we vary the duration of missing cellular locations $t$ from $20$ seconds to $100$ seconds by sampling cellular locations from the raw cellular trajectories, and the accuracy of the transition probability can be evaluated by the {\em probability difference ratio}  for transitive probability ($PDR^{TP}$): $\frac{p_{max}(ef_{i+1}\in \mathcal{R}_{cl_{i+1}}|ef_i)-p_{a}(ef_{a}|ef_i)}{p_{max}(ef_{i+1}\in \mathcal{R}_{cl_{i+1}}|ef_i)}$, where $p_{max}(ef_{i+1}\in \mathcal{R}_{cl_{i+1}}$ $|ef_i)$ denotes the maximum transition probability for the transition from current edge fragment $ef_i$ to an edge fragment in $\mathcal{R}_{cl_{i+1}}$, and $p_{a}(ef_{a}|ef_i)$ is the transition probability for the actually happened transition.
	It is clear that a smaller probability difference ratio is preferred.
	Figure~\ref{fig:trans-prob:vt} shows the average probability difference ratios by varying time gaps (i.e., missing location lengths).
	We can see that the proposed adaptive transition probability maintains a low probability difference ratio as $t$ grows, indicating it gives a higher probability to the actually traveled edge fragment pair.
	In contrast, the PF, which does not consider the traffic conditions, deteriorates as $t$ grows, suggesting it is incapable of depicting the long-distance movements. 

	\textbf{Compact $\mathcal{R}_{cl}$ Derivation:} We study the pruning power of two pruning techniques in varying trajectories. The pruning power is evaluated by the ratio of the derived $\mathcal{R}_{cl}^\star$ count to the number of cellular locations in $\mathcal{T}$, called {\em compact ratio} $r$.
	A large $r$ is preferred. However, given the high spatial error, it is difficult to find many most compact edge fragment sets.
	As shown in Figure~\ref{fig:two-f:pf}, the $r$ values oscillates over various trajectories. On average, $r$ values produced by pairwise pruning can only reach around $0.04$.
	Although a limited number of the most compact edge fragment sets are found by pairwise pruning, sequence pruning is capable of boosting $r$ values since a discovered $\mathcal{R}^\star_{cl}$ propagates in the sequence pruning.
	In Figure~\ref{fig:two-f:sf}, the average $r$ value after sequence pruning is around $0.15$, indicating $15~\%$ of all cellular locations are firmly aligned to a particular transportation edge after applying two pruning techniques.
	
	\textbf{Online} $\mathcal{D}_e$ \textbf{Learning}: We evaluate the online learning method by comparing the arithmetic mean $\widetilde{t_e}$ of travel times from the narrowed time range $R$ with $\widetilde{t}$ derived from all available travel times. 
	Such a comparison is done by {\em accuracy loss ratio}: $\frac{|\widetilde{t_e}-\widetilde{t}|}{\widetilde{t}}$. 
	Figure~\ref{fig:de-learn:ti} shows that with the growth of $\delta$ narrowing procedure gradually reduces the accuracy loss ratio between the online learning and the batch learning based on all travel times, however, the reduction is not substantial for $\delta\in [0.05,0.15]$.
	Figure~\ref{fig:de-learn:varyd} shows the results from varying $\epsilon$.
	We can see that with the growth of $\epsilon$, the accuracy loss ratio reduces.
	$\epsilon=2$ brings a clear accuracy loss ratio reduction.
	Therefore, by default, we choose $\delta=0.05$ and $\epsilon=2$, which brings us the significantly narrowed time range and the minor accuracy loss ratio.

	\textbf{Varying $|\mathcal{W}|$}.
	We study the impact of $|\mathcal{W}|$ of DTNC.
	As shown in Figure~\ref{fig:varying-w}, the cleansed cellular locations gradually achieve higher accuracy as $|\mathcal{W}|$ increases.
	This is because the inference algorithm described in Algorithm~\ref{alg:viterbi-hsmm-mv} computes the most probable edge sequence based on more cellular locations, which achieves a more reliable result.
	Nevertheless, the growth of the accuracy slows down after $|\mathcal{W}|=70$.
	Based on our discussion with StarHub, this may be due to the switch of  servicing cellular towers, that result in higher spatial errors.
	Thus, in our experiments, $|\mathcal{W}|$ is set as $70$ by default.

	\subsection{Additional Experiments}
	
	\begin{figure}
		\subfigure[Varying Particle Numbers]{
			\label{fig:varying-A-rt}
		\includegraphics[width=0.47\columnwidth,height=3cm]{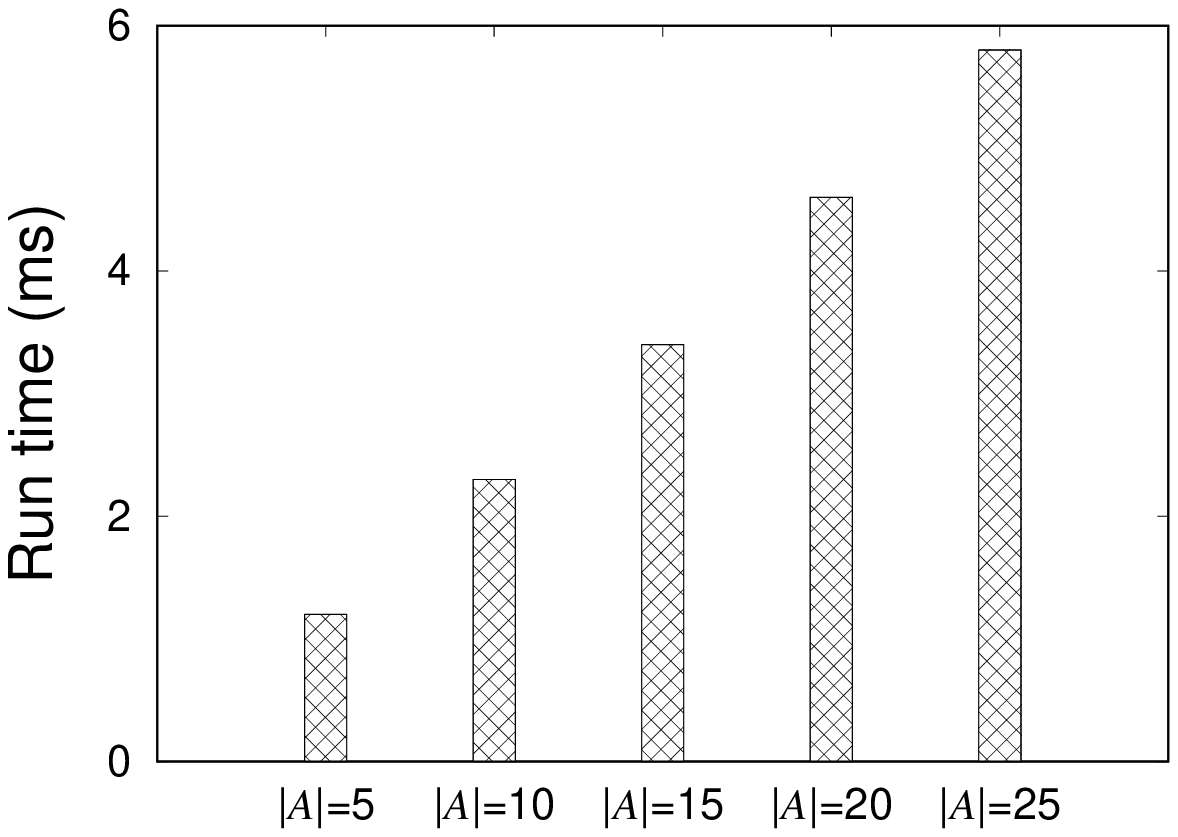}
		}
		\subfigure[Varying Diffusion Policies]{
			\label{fig:varying-dps}
			\includegraphics[width=0.48\columnwidth,height=3cm]{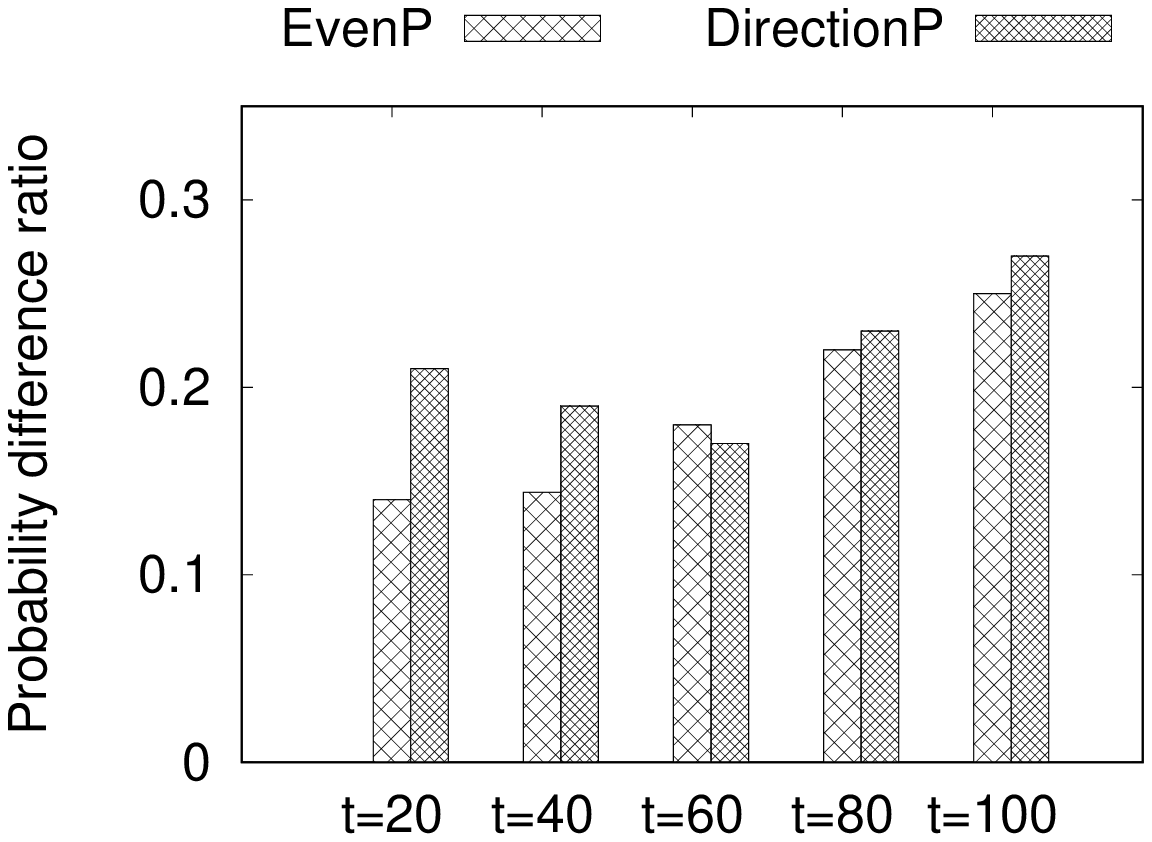}}
		\vspace{-10pt}
		\caption{Particle Diffusion Evaluation}
		\label{fig:particle-filter}
		\vspace{-12pt}
	\end{figure}

	\textbf{Varying Particle Size}.
	Incorporating more particles helps improve the accuracy of the approximated adaptive transition probability.
	However, the run time also grows due to the increase of more particles.
	In this case, we study the effect of varying particle sizes and perform trade-off analysis.
	The evaluation results are shown in Figure
	~\ref{fig:varying-A-rt}.
	%
	which shows the {\em probability difference ratio}s (same as the one used in adaptive transition probability evaluation) w.r.t. varying particle number $|\mathcal{A}|$.
	We can see that the {\em probability difference ratio} slows down after $|\mathcal{A}|=15$ for our dataset $D_1$.
	And as shown in Figure~\ref{fig:varying-A-rt}, the run time linearly increases as $|\mathcal{A}|$ grows.
	Therefore, by default, $|\mathcal{A}|=15$ is used for $D_1$, which takes moderate run time and achieves relatively low {\em probability difference ratio}.
	
	\textbf{Varying Diffusion Policies}. 
	By setting $m_1=\cdots=m_{|\mathcal{R}_{cl_k}|}=\frac{1}{|\mathcal{R}_{cl_k}|}$, we assume that each edge fragment has the same chance to be traveled. 
	We call such parameter setting as {\em even policy}, denoted by {\em evenP}.
	
	Alternatively, we can heuristically assign the $m$ values in accordance with their directional coherences.
	We dub such heuristic setting as {\em directional policy}, denoted by {\em directionP}.
	Concretely, {\em directionP} gives the edge fragments owning similar directions with the subsequent cellular locations higher probability.
	In this light, the edge fragments are categorized into {\em positive}, {\em negative} and {\em neutral} directions,
	where {\em positive direction} means $cos(ef,cl_k cl_{k+1})>0$, {\em negative direction} means $cos(ef, cl_k cl_{k+1})<0$, and {\em neutral direction} means $cos(ef, cl_k cl_{k+1})=0$.
	For a {\em positive-direction} edge fragment, its $m=\frac{2\alpha}{|\mathcal{R}_{cl_k}|}$; for a {\em neutral-direction} one, its $m=\frac{1}{|\mathcal{R}_{cl_k}|}$; and for a {\em negative-direction} one, its $m=\frac{0.5\alpha}{|\mathcal{R}_{cl_k}|}$.
	According to $\sum_{ef\in \mathcal{R}_{cl_k}}=1$, $\alpha$ can be computed.
	
	Figure~\ref{fig:varying-dps} plots the evaluation results w.r.t. {\em evenP} and {\em directionP} when the time gap $t$ varies.
	Here, {\em probability difference ratio} is same as the one used in the adaptive transition probability evaluation.
	Not surprisingly, {\em DirectionP} performs better when $t$ is not large.
	This validates the effectiveness of the heuristics.
	Nevertheless, as $t$ grows, the difference between {\em evenP} and {\em DirectionP} is not significant for $D_1$, because the direction information is not so helpful when $t$ is large.
	Therefore, by default, we adopt {\em EvenP} as the particle diffusion policy for it is simple yet effective.

	\begin{figure}
		\includegraphics[width=0.95\columnwidth]{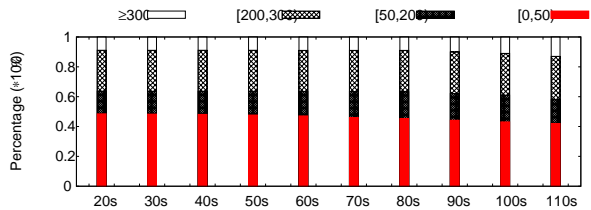}
		\caption{Varying Missing Location Length}
		\label{fig:eff-acc-varying-time}
	\end{figure}
	
	\textbf{Varying Missing Location Length}
		As arbitrary missing location is the other challenge, 
		we sample cellular locations from the raw cellular trajectories to construct trajectories with various time gaps for missing locations.
		We call the time gap between two consecutive cellular locations as missing location length.
		We apply DTNC to the constructed trajectories to study its {\em Euclidean deviation} distributions.
		Note that we have put all the constructed trajectories (from $\mathcal{T}_1$ to $\mathcal{T}_{10}$) together in this experiment.
		
		Figure~\ref{fig:eff-acc-varying-time} shows the results of the {\em Euclidean deviation}s, where the missing location length varies from $20$~seconds to $110$~seconds.
		The results are grouped by time gaps.
		%
		
		As the time gap grows, DTNC maintains a relatively stable distribution of {\em Euclidean deviation}. 
		The high-quality location (with spatial error below $50$~meters) rate of DTNC gradually decreases from $47\%$ to $41\%$, suggesting it is good at dealing with the missing values.
		%
		%
		
		The reasons why DTNC is not sensitive to the missing locations length are two-fold.
		First, DTNC accurately simulates the movement between two cellular locations, based on the continuously updated edge travel time distributions.
		%
		%
		Thus, it is capable of differentiating paths in terms of the traffic-aware simulations. 
		Second, 
		%
		only the particles moving along the most probable edge sequence are adopted to infer physical locations.
		Hence, it is insensitive to the missing locations, and is able to produce accurate results.

	\section{Conclusion}
	\label{sec:conclusion}
	
	%
	In this paper, we propose a new data cleansing framework called \underline{D}ynamic \underline{T}ransportation \underline{N}etwork based \underline{C}leansing (DTNC), which produces cleansed cellular trajectories with accurate locations. 
	DTNC utilizes real-time traffic information maintained in a dynamic transportation network to derive robust emission probabilities, adaptive transition probabilities as input to the proposed object motion model (i.e., TT-HsMM) in order to infer the most probable edge sequence for data cleansing.
	
	An evaluation with real-world cellular data offers insight into the design of DTNC and confirms its effectiveness and efficiency.
	In particular, after being processed by DTNC, the spatial errors in $40\%$ of cellular locations are reduced to below $50$~meters. 
	Further, DTNC significantly outperforms its potential state-of-the-art rivals in CTC, including  PF, HMM, OHMM, STRS, SnapNet, and CTrack.
	
	{
		\small
		\bibliographystyle{abbrv}
		\bibliography{sigproc-short} 

\begin{thebibliography}{10}

\bibitem{itu-int}
The {ITU ICT} facts and figures.
\newblock
  \url{http://www.itu.int/en/ITU-D/Statistics/Documents/facts/ICTFactsFigures2015.pdf}.
\newblock Accessed: 2016-07-07.

\bibitem{blondel2012data}
V.~D. Blondel, M.~Esch, C.~Chan, F.~Cl{\'e}rot, P.~Deville, E.~Huens,
  F.~Morlot, Z.~Smoreda, and C.~Ziemlicki.
\newblock Data for development: the d4d challenge on mobile phone data.
\newblock {\em arXiv preprint arXiv:1210.0137}, 2012.

\bibitem{calabrese2011estimating}
F.~Calabrese, G.~Di~Lorenzo, L.~Liu, and C.~Ratti.
\newblock Estimating origin-destination flows using mobile phone location data.
\newblock {\em IEEE Pervasive Computing}, 10(4), 2011.

\bibitem{calabrese2013understanding}
F.~Calabrese, M.~Diao, G.~Di~Lorenzo, J.~Ferreira, and C.~Ratti.
\newblock Understanding individual mobility patterns from urban sensing data: A
  mobile phone trace example.
\newblock {\em Transportation research part C: emerging technologies},
  26:301--313, 2013.

\bibitem{chang2005multiobjective}
T.-S. Chang, L.~K. Nozick, and M.~A. Turnquist.
\newblock Multiobjective path finding in stochastic dynamic networks, with
  application to routing hazardous materials shipments.
\newblock {\em Transportation science}, 39(3):383--399, 2005.

\bibitem{chen2015fusion}
Z.~Chen, H.~Zou, H.~Jiang, Q.~Zhu, Y.~C. Soh, and L.~Xie.
\newblock Fusion of wifi, smartphone sensors and landmarks using the kalman
  filter for indoor localization.
\newblock {\em Sensors}, 15(1):715--732, 2015.

\bibitem{cong2002hybrid}
L.~Cong and W.~Zhuang.
\newblock Hybrid tdoa/aoa mobile user location for wideband cdma cellular
  systems.
\newblock {\em IEEE Trans. Wireless Commun.}, 1(3):439--447, 2002.

\bibitem{dai2016path}
J.~Dai, B.~Yang, C.~Guo, C.~S. Jensen, and J.~Hu.
\newblock Path cost distribution estimation using trajectory data.
\newblock {\em Proceedings of the VLDB Endowment}, 10(3):85--96, 2016.

\bibitem{goh2012online}
C.~Y. Goh, J.~Dauwels, N.~Mitrovic, M.~Asif, A.~Oran, and P.~Jaillet.
\newblock Online map-matching based on hidden markov model for real-time
  traffic sensing applications.
\newblock In {\em ITSC}, pages 776--781, 2012.

\bibitem{gustafsson2002particle}
F.~Gustafsson, F.~Gunnarsson, N.~Bergman, U.~Forssell, J.~Jansson, R.~Karlsson,
  and P.-J. Nordlund.
\newblock Particle filters for positioning, navigation, and tracking.
\newblock {\em IEEE Transactions on signal processing}, 50(2):425--437, 2002.

\bibitem{hoeffding1963probability}
W.~Hoeffding.
\newblock Probability inequalities for sums of bounded random variables.
\newblock {\em Journal of the American statistical association},
  58(301):13--30, 1963.

\bibitem{hu2017if}
G.~Hu, J.~Shao, F.~Liu, Y.~Wang, and H.~T. Shen.
\newblock If-matching: Towards accurate map-matching with information fusion.
\newblock {\em TKDE}, 29(1):114--127, 2017.

\bibitem{hua2010probabilistic}
M.~Hua and J.~Pei.
\newblock Probabilistic path queries in road networks: traffic uncertainty
  aware path selection.
\newblock In {\em EDBT}, pages 347--358. ACM, 2010.

\bibitem{kang2015smartpdr}
W.~Kang and Y.~Han.
\newblock Smartpdr: Smartphone-based pedestrian dead reckoning for indoor
  localization.
\newblock {\em IEEE Sensors journal}, 15(5):2906--2916, 2015.

\bibitem{kempinska2016probabilistic}
K.~Kempinska, T.~Davies, and J.~Shawe-Taylor.
\newblock Probabilistic map-matching using particle filters.
\newblock {\em arXiv preprint arXiv:1611.09706}, 2016.

\bibitem{krumm2007map}
J.~Krumm, E.~Horvitz, and J.~Letchner.
\newblock Map matching with travel time constraints.
\newblock Technical report, 2007.

\bibitem{lian2014trip}
X.~Lian and L.~Chen.
\newblock Trip planner over probabilistic time-dependent road networks.
\newblock {\em IEEE Transactions on Knowledge and Data Engineering},
  26(8):2058--2071, 2014.

\bibitem{lou2009map}
Y.~Lou, C.~Zhang, Y.~Zheng, X.~Xie, W.~Wang, and Y.~Huang.
\newblock Map-matching for low-sampling-rate gps trajectories.
\newblock In {\em SIGSPATIAL}, pages 352--361, 2009.

\bibitem{mohamed2016accurate}
R.~Mohamed, H.~Aly, and M.~Youssef.
\newblock Accurate real-time map matching for challenging environments.
\newblock {\em IEEE Transactions on Intelligent Transportation Systems}, 2016.

\bibitem{newson2009hidden}
P.~Newson and J.~Krumm.
\newblock Hidden markov map matching through noise and sparseness.
\newblock In {\em SIGSPATIAL}, 2009.

\bibitem{pfoser1999capturing}
D.~Pfoser and C.~S. Jensen.
\newblock Capturing the uncertainty of moving-object representations.
\newblock In {\em International Symposium on Spatial Databases}, pages
  111--131. Springer, 1999.

\bibitem{phithakkitnukoon2010activity}
S.~Phithakkitnukoon, T.~Horanont, G.~Di~Lorenzo, R.~Shibasaki, and C.~Ratti.
\newblock Activity-aware map: Identifying human daily activity pattern using
  mobile phone data.
\newblock In {\em International Workshop on Human Behavior Understanding},
  pages 14--25. Springer, 2010.

\bibitem{raymond2012map}
R.~Raymond, T.~Morimura, T.~Osogami, and N.~Hirosue.
\newblock Map matching with hidden markov model on sampled road network.
\newblock In {\em ICPR}, pages 2242--2245, 2012.

\bibitem{steenbruggen2015data}
J.~Steenbruggen, E.~Tranos, and P.~Nijkamp.
\newblock Data from mobile phone operators: A tool for smarter cities?
\newblock {\em Telecommunications Policy}, 39(3):335--346, 2015.

\bibitem{su2015calibrating}
H.~Su, K.~Zheng, J.~Huang, H.~Wang, and X.~Zhou.
\newblock Calibrating trajectory data for spatio-temporal similarity analysis.
\newblock {\em The VLDB Journal}, 24(1):93--116, 2015.

\bibitem{thiagarajan2011accurate}
A.~Thiagarajan, L.~Ravindranath, H.~Balakrishnan, S.~Madden, L.~Girod, et~al.
\newblock Accurate, low-energy trajectory mapping for mobile devices.
\newblock In {\em NSDI}, 2011.

\bibitem{wei2012fast}
H.~Wei, Y.~Wang, G.~Forman, Y.~Zhu, and H.~Guan.
\newblock Fast viterbi map matching with tunable weight functions.
\newblock In {\em SIGSPATIAL}, pages 613--616, 2012.

\bibitem{DBLP:conf/kdd/WuMSZZCW16}
H.~Wu, J.~Mao, W.~Sun, B.~Zheng, H.~Zhang, Z.~Chen, and W.~Wang.
\newblock Probabilistic robust route recovery with spatio-temporal dynamics.
\newblock In {\em SIGKDD}, 2016.

\bibitem{wu2016only}
H.~Wu, W.~Sun, and B.~Zheng.
\newblock Is only one gps position sufficient to locate you to the road network
  accurately?
\newblock In {\em UbiComp}, pages 740--751, 2016.

\bibitem{yang2014stochastic}
B.~Yang, C.~Guo, C.~S. Jensen, M.~Kaul, and S.~Shang.
\newblock Stochastic skyline route planning under time-varying uncertainty.
\newblock In {\em ICDE}, pages 136--147, 2014.

\bibitem{yin2016general}
Y.~Yin, R.~R. Shah, and R.~Zimmermann.
\newblock A general feature-based map matching framework with trajectory
  simplification.
\newblock In {\em IWGS}, 2016.

\bibitem{DBLP:journals/sigpro/YuK03}
S.~Yu and H.~Kobayashi.
\newblock A hidden semi-markov model with missing data and multiple observation
  sequences for mobility tracking.
\newblock {\em Signal Processing}, 83(2):235--250, 2003.

\bibitem{zheng2012reducing}
K.~Zheng, Y.~Zheng, X.~Xie, and X.~Zhou.
\newblock Reducing uncertainty of low-sampling-rate trajectories.
\newblock In {\em ICDE}, pages 1144--1155, 2012.

\end{thebibliography}
	}
		
\end{document}